\documentclass[preprint2]{aastex}
\usepackage{longtable}
\usepackage{lscape}
\usepackage{ulem}
\usepackage{array}

\newcommand{\mysplit}[1]{%
  \begin{tabular}[t]{@{}c@{}}   
    #1
  \end{tabular}}

\def\etal{\it et~al.}

\shorttitle{}
\shortauthors{Skrzypczak \etal}

\begin{document}

\title{Meterwavelength Single-pulse Polarimetric Emission Survey IV: 
The Period Dependence of Component Widths of Pulsars.} 

\author{Anna Skrzypczak\altaffilmark{1}, Rahul Basu\altaffilmark{2,1}, Dipanjan Mitra\altaffilmark{3,4,1}, George I. Melikidze\altaffilmark{1,5}, Krzysztof Maciesiak\altaffilmark{1,7}, Olga Koralewska\altaffilmark{1}, Alexandros Filothodoros\altaffilmark{6}} 

\altaffiltext{1}{Janusz Gil Institute of Astronomy, University of Zielona G\'ora, ul. Szafrana 2, 65-516 Zielona G\'ora, Poland}
\altaffiltext{2}{Inter-University Centre for Astronomy and Astrophysics, Pune 411007, India}
\altaffiltext{3}{National Centre for Radio Astrophysics, Ganeshkhind, Pune 411 007, India}
\altaffiltext{4}{Physics Department, University of Vermont, Burlington VT 05405}
\altaffiltext{5}{Abastumani Astrophysical Observatory, Ilia State University, 3-5 Cholokashvili Ave., Tbilisi, 0160, Georgia}
\altaffiltext{6}{Institute of Physics, University of Zielona G\'ora, ul. Prof. Szafrana 4a, 65-516 Zielona G\'ora, Poland}
\altaffiltext{7}{Botswana International University of Science and Technology, Private Bag 16, Palapye, Botswana}
\email{hera$\_$ania@o2.pl}

\begin{abstract}
\noindent
The core component width in normal pulsars, with periods ($P$) $>$ 0.1 seconds,
measured at the half-power point at 1 GHz has a lower boundary line (LBL) which
closely follows the $P^{-0.5}$ scaling relation. This result is of fundamental 
importance for understanding the emission process and requires extended studies
over a wider frequency range. In this paper we have carried out a detailed 
study of the profile component widths of 123 normal pulsars observed in the 
Meterwavelength Single-pulse Polarimetric Emission Survey at 333 and 618 MHz. 
The components in the pulse profile were separated into core and conal classes.
We found that at both frequencies the core as well as the conal component 
widths versus period had a LBL which followed the $P^{-0.5}$ relation with a 
similar lower boundary. The radio emission in normal pulsars have been 
observationally shown to arise from a narrow range of heights around a few 
hundred kilometers above the stellar surface. In the past the $P^{-0.5}$ 
relation has been considered as evidence for emission arising from last open 
dipolar magnetic field lines. We show that the $P^{-0.5}$ dependence only holds
if the trailing and leading half-power points of the component are associated 
with the last open field line. In such a scenario we do not find any physical 
motivation which can explain the $P^{-0.5}$ dependence for both core and conal 
components as evidence for dipolar geometry in normal pulsars. We believe the 
period dependence is a result of an yet unexplained physical phenomenon. 
\end{abstract}


\section{\large Introduction} \label{sec:intro}
\noindent
The coherent radio emission from pulsars are believed to arise due to growth of
instabilities in relativistic strongly magnetized plasma outflowing along open 
dipolar magnetic field lines well inside the light cylinder, however the 
underlying physical mechanism that excites the coherent radio emission is still
unidentified \citep[see e.g.][]{mic82,bes93,ass98,mel00,mel17,mit17}. The
radio pulsar population can be roughly divided into two groups based on their 
periods, millisecond pulsars with periods much less then 100 milliseconds and 
normal pulsars with periods longer than this value. The presence of dipolar 
magnetic fields in the emission region is motivated by the widths of the 
profile and specific components which exhibit a dependence on the period ($P$).
In this work we take a critical look into this argument utilizing the measured 
profiles of 123 normal pulsars with periods greater than 0.1 seconds observed 
in the Meterwavelength Single-pulse Polarimetric Emission Survey \citep[MSPES,
][]{mit16b,bas16}. Only for pulsars with periods larger than 0.1 second will an 
emission region that is located roughly 500 km above the neutron star \citep{
bla91,von97,kij97,mit04,wel08,krz09,mit11} be largely unaffected by strong 
field line distortions \citep{dyk08}.

The periodic radio emission in normal pulsars has typical duty cycles of 
$\sim$ 10\% and the single pulses often consist of smaller structures called 
subpulses. In some cases the subpulses appear at the same phase within the 
pulse window while in others they jitter or systematically move in phase. When 
the single pulses are averaged for a few thousand periods a stable integrated 
profile is formed which is composed of one or more distinct gaussian shaped 
components. These components are formed due to the averaging of subpulses but 
their properties like width and location might differ from the individual 
subpulses depending on the subpulse dynamics. The physical origin of the 
subpulses in single pulses are still unknown. The most notable radio emission 
model that predict the presence of subpulses is that of \citet{rud75} where the
subpulses are associated with radiation generated from isolated plasma columns.
The plasma columns are generated due to sparking discharges in an inner 
accelerating region characterised by high electric fields just above the polar 
cap. The formation of subpulses have also been speculated to arise due to 
non-radial oscillations in neutron star \citep{cle04} or due to development of 
instability in the outflowing plasma \citep{fun06}. This motivates the study of
average subpulse properties in a large number of pulsars to constrain the 
formation of these emission structures.

The number of components in the profiles are seen to vary in the pulsar 
population. The full widths of the profiles usually exhibit a frequency 
evolution where the widths are seen to decrease with increasing frequency and 
the number of components often change with frequency in a systematic manner. 
The profiles are highly polarized and the linear polarization position angle 
(PPA) executes a S-shaped traverse across them. The rotating vector model 
\citep[RVM,][]{rad69} is used to explain the shape of the PPA traverse where 
beamed emission arises from regions of diverging, but not necessarily dipolar 
magnetic field lines with the PPA traverse corresponding to the change
in the field line planes. In addition, the PPA traverse is also affected by the
observer's line of sight cut through the pulsar emission beam. If the diverging
magnetic field lines are ascribed to a star centered global dipolar magnet, the
RVM is expressed as a function of the dipolar geometrical angles, viz. the 
angle between the rotation axis and the dipole magnetic axis, $\alpha$, and the
angle between the dipole magnetic axis and the observer's line of sight 
$\beta$. The RVM has been shown to fit the PPA traverse in normal pulsars
very well indicating  the emission to originate from dipolar magnetic fields
\citep{eve01,mit04}. On the other hand these fits are very often unsuitable for
estimates of the angles $\alpha$ and $\beta$ when data exhibit large error bars
on PPA. The fit residuals become strongly correlated and good constraints for 
an unique solution are absent. However, the steepest gradient (SG) point of the
PPA which lies in the plane containing the rotation and dipole magnetic axis 
can often be used to provide the necessary additional constraint 
\begin{equation}
R_{ppa} = \sin(\alpha)/\sin(\beta). 
\label{eq1}
\end{equation}
where $R_{ppa}$ corresponds to the slope of the PPA at SG, for a more accurate 
determination of $\alpha$ and $\beta$.

A number of studies in the literature details the interplay between the 
magnetospheric plasma with the field configuration, including the polar cap 
structure \citep{bai10}, the structure of the large scale magnetic field and 
plasma \citep{phi14,phi15}, the location of discharge regions and plasma 
sources \citep{che14}, the dynamics of discharge regions \citep{sza15,tim13} 
and propagation effects \citep{pet00,hak14}. Additionally, the outflowing 
plasma in the rotating magnetosphere can also lead to effects like abberation 
and retardation (A/R hereafter), magnetic field sweepback, etc. These can 
potentially distort the shape of the PPA traverse, yet the RVM corresponding to
a relatively simplified model of empty pulsar magnetosphere are excellent fits 
to the PPA in normal pulsars \citep{eve01,mit04}. Several studies have shown 
that if the radio emission detaches from the magnetosphere at heights of 
$h_{em} < 0.1R_{LC}$, where $R_{LC}$ = $pc/2\pi$ is the light cylinder radius, 
the A/R effects are the only discernible observable distortion affecting the 
PPA traverse \citep{bla91,hib01,dyk08,cra12,kum12a,kum12b,kum13}. In such a 
scenario if the radio emission detaches at a constant height across the pulsar 
profile, the A/R effect can be approximated as a positive shift in longitude, 
$\Delta \phi$, between the center of the total intensity profile and the SG 
point of the PPA traverse, which is linearly dependent on $h_{em}$. This can in
turn be used to estimate the emission heights as $h_{em} = (c/4) (\Delta 
\phi/360\degr) P$ km. Using the above relation a number of observations have 
found the radio emission in normal pulsars to detach from the magnetosphere 
around $h_{em}\sim$ 100--1000 km, which is well below 0.1$R_{LC}$ \citep{bla91,
von97,mit04,wel08,mit11}. It should be noted that the excellent fits of the PPA
traverse with the RVM indicates that the emission region across the pulse 
profile originate from similar heights, since significant changes in emission 
height can distort the PPA traverse (see for e.g. \citet{mit04b}). The radio 
emission is excited in the plasma at a certain height $h_{g}$, where $h_{g} < 
h_{em}$, and then it propagates in the plasma and eventually detaches at 
$h_{em}$. The plasma properties in the propagation region can influence the 
properties of the radio emission and several conflicting models exist in the 
literature which either suggest that the radiation is unaffected by propagation
\citep[][]{mit09,mel14} or that the radiation is modified due to the 
propagation effect \citep[][]{pet00,hak14}. However, the observational 
constraints from pulsar radiation can still be used to study the magnetospheric
structure at the detachment point $h_{em}$. At these heights the estimates of 
the locus of the open dipolar field lines suggest the emission region to be 
close to a circular patch \citep{are02,dyk04}. This motivates the idea of the 
emission components in the pulsar profile of normal pulsars to originate from 
an emission beam radiating approximately at a constant height and the angular 
dimensions of the emission component and beams can be constrained by spherical 
geometry.

The components can be separated into core and conal types which exhibit 
distinct frequency evolution as well as different polarization characteristics,
with the core always centrally located within the profile. 
The emission beam in pulsars has been conceptualized as a circularly symmetric
structure consisting of a central core component surrounded by concentric rings
of conal components \citep[][R90 and R93 hereafter]{ran90,ran93a,ran93b}. 
The principal justification for the above picture is provided by the different 
profile shapes in the pulsar population which can be categorized into five 
distinct classes. The five component profiles consisting of a central core 
component flanked by two pairs of inner and outer conal outriders are 
classified as Multiple ({\bf M}) class. There are two classes of three 
component profiles namely core Triple ({\bf T}) which has a central core 
component along with one pair of conal outriders and the conal Triple 
({\bf $_c$T}) where all the three components are conal. The two component 
profiles correspond to conal Double ({\bf D}) class. Finally, the single 
component profiles are classified either as Core Single ({\bf S$_t$}) or
the Conal Single ({\bf S$_d$}) classes. It is believed that the different 
profile classes arise due to different line of sight cuts through the emission 
beam and is therefore dependent on the pulsar geometry, with the beam radius 
($\rho^{\nu}$) given as \citep{gil84} :
\begin{equation}
\sin^2(\rho^{\nu}/2) = \sin(\alpha)\sin(\alpha+\beta)\sin^2(W^{\nu}/4)
+\sin^2(\beta/2)
\label{eq2}
\end{equation}
where $W^{\nu}$ is the estimated profile width at frequency $\nu$. Further, by 
assuming $W^{\nu}$ to be bound by the last open dipolar field lines, an 
estimate of the emission height ($h^{\nu}$) above the neutron star polar cap 
can be computed as: 
\begin{equation}
h^{\nu} = 10 P \left(\frac{\rho^{\nu}}{1.23\degr}\right)^2 {\rm km}
\label{eq3}
\end{equation}
where the radius of the opening angle of the polar cap at the stellar surface
($R$ = 10 km) corresponds to $1.23\degr$. As discussed above the pulsar 
geometry, angles $\alpha$ and $\beta$, cannot be constrained using the RVM. 
Hence, $h^{\nu}$ cannot be calculated with any certainty using equations 
(\ref{eq2}) and (\ref{eq3}). 

R90 established a dependence of the core component widths with pulsar 
period which enabled an alternative scheme for estimating the pulsar geometry. 
The widths measured at 1~GHz $W_{core}^{1 {\rm GHz}}$ (estimated at 50\% of 
peak intensity) showed the presence of a lower boundary line (LBL) scaling as 
$2.45\degr P^{-0.5}$. R90 further established that core widths for several 
inter-pulsars (where $\alpha \sim 90\degr$) lay along the LBL. It should be 
noted that pulsars with core components have the line of sight passing 
centrally through the emission beam and hence their $\beta$ is small. 
Incidentally, the boundary value $2.45\degr$ is very close to the diameter of 
the dipolar opening angle at the stellar surface. It was postulated that 
$W_{core}^{1 {\rm GHz}}$ encompasses the pulsar polar-cap bounded by last open 
dipolar field lines. The pulsars with the measured widths above the line 
corresponds to non-orthogonal rotators with $\sin\alpha < 1$. This allowed an 
independent estimation of the angle $\alpha$ from $W_{core}^{1 {\rm GHz}}$ 
using the relation:
\begin{equation}
W_{core}^{1 {\rm GHz}} = 2.45 P^{-0.5}/\sin\alpha 
\label{eq4}
\end{equation}

R93 expanded the above idea to estimate the geometry in a large number 
of pulsars with core components. The conal separation in profile classes 
{\bf T} and {\bf M}, with prominent core components, were also measured. It was 
proposed that the inner and outer cones originate at different heights and 
encompass the entire open field line region. The half-power points of inner and
outer conal pairs ($W_{in,out}^{1 {\rm GHz}}$) also enabled the estimation of 
the radius of the opening angle ($\rho_{in,out}^{1 {\rm GHz}}$) for the conal 
rings as :
\begin{equation}
\rho_{in}^{1 {\rm GHz}}=4.3\degr P^{-0.5}; \rho_{out}^{1 {\rm GHz}}=5.8\degr P^{-0.5}
\label{eq5}
\end{equation}
R93 estimated the inner and outer cones to arise from heights 
(eq.\ref{eq3}) of roughly 130 km and 220 km respectively. The method of
analysing profile widths and components by R90, R93 has an underlying 
assumption that all relevant widths arise from the entire open field line 
region with an emphasis on the core, which is believed to fill up the entire 
polar cap, the inner and outer conal pairs, once again representing the entire 
open field line regions, etc. There are no provisions for interpreting the 
widths of the individual components in these schemes.

The above picture of the radio emission is difficult to reconcile with recent 
observations where emission heights of the core and conal components is emitted
from similar heights in the same  pulsar. It has now been shown in many {\bf M}
and {\bf T} class pulsars that the core and conal emission originates from 
similar heights \citep{mit07,mit11,smi13,mit16a}. In this context the 
$P^{-0.5}$ scaling relation between $W_{core}^{\nu}$ and $\rho^{\nu}$ needs 
careful consideration. If we assume $\beta$ to be small and $\alpha >> \beta$ 
(which is justified for central cuts with core emission), the radius of the 
emission beam in eq.(\ref{eq2}) can be approximated as :
\begin{equation}
\rho^{1 {\rm GHz}} = \sqrt{4\sin^2(\alpha)\sin^2(W^{1 {\rm GHz}}/4)+\beta^2}.
\label{eq6}
\end{equation}
Now substituting sin($\alpha$) in terms of core width (eq.\ref{eq4}, 
which introduces the $P^{-0.5}$ dependence) and $\beta$ in terms of 
sin($\alpha$) and $R_{ppa}$ (eq.\ref{eq1}) the emission beam is estimated as:
\begin{equation}
\rho^{1 {\rm GHz}} = 2.45 P^{-0.5} F
\label{eq7}
\end{equation}
where the factor $F$ is given by
\begin{equation}
F = \sqrt{\left(\frac{4\sin^2(W^{1 {\rm GHz}}/4)}{(W_{core}^{1 {\rm GHz}})^2}+\frac{1}{({R_{ppa} W_{core}^{1 {\rm GHz}}})^2}\right)}
\label{eq8}
\end{equation}
As is clear from the above exercise the $P^{-0.5}$ dependence in 
$\rho^{1 {\rm GHz}}$ is transferred from the $P^{-0.5}$ dependence of the 
core width and is only preserved if the factor $F$ does not have any period
dependence. 

To further investigate this in {\bf M} and {\bf T} class profiles with 
prominent core and conal components we recall that $R_{ppa}$ is usually large 
and hence the second term in eq.(\ref{eq8}) can be ignored. Now if we assume 
that the total width $W^{1 {\rm GHz}}$ to be small, $F$ can be approximated as
\begin{equation}
F \sim W^{1 {\rm GHz}}/2W_{core}^{1 {\rm GHz}}. 
\label{eq9}
\end{equation}
The total width $W^{1 {\rm GHz}}$ can be separated into individual components:
\begin{equation}
W^{\nu} = W_{core}^{\nu}+\sum_i W_{cone,i}^{\nu}+\sum_j \delta W_{j}^{\nu}
\label{eq10}
\end{equation}
where $W_{cone,i}^{\nu}$ is the width of the $i^{th}$ conal component, 
$\delta W_{j}^{\nu}$ is the $j^{th}$ separation between adjacent components 
with the summation extending over all conal components as well as the 
respective separation between adjacent components. Now, For the factor $F$ to 
be period independent $W^{1 {\rm GHz}}$ should also show a $P^{-0.5}$ 
dependence. This further implies that along with $W_{core}^{\nu}$, the 
$W_{cone,i}^{1 {\rm GHz}}$ and $\delta W_{j}^{1 {\rm GHz}}$ should also have 
$P^{-0.5}$ dependence.

To the best of our knowledge no such study exists in the literature connecting 
the conal components widths with the pulsar period. Only some hints of this 
effect have been discussed in \citet{mac11a,mac11b,mac12} and \citet{mit16b}. In
these works a distribution of total half-power width of all available profiles 
with period found a LBL which scaled as $P^{-0.5}$. Since no particular profile 
class was selected to obtain this relation, a LBL might exist for both core and
conal widths following a similar $P^{-0.5}$ relation. In this paper, we use the
MSPES data to carry out a detailed study of the profile component widths and 
investigate their dependence on period. Furthermore, we explore the 
implications of the period dependence of component widths on the dipolar 
geometry.

\section{\large Component Width Analysis} \label{sec:obs}
\noindent
The principal analysis involved estimating the relevant widths of the 
components and the separation between adjacent components. Generally, the 
widths can be directly measured in the integrated profiles where the components
are clearly distinguished. But in some pulsars, due to phenomenon like subpulse
drifting, the subpulses are seen to systematically drift in phase across the 
pulse window. It is also possible that in the same pulsar different components 
might be associated with different subpulse dynamics resulting in different 
widths. In such cases a more accurate estimate of the emission properties are 
possible by correcting for the subpulse motion across the pulse window and form
an average component from selected single pulses.

We used the MSPES data at 333 and 618 MHz for 123 pulsar with high quality 
single pulses for this purpose \citep{mit16b}. Specialized techniques were 
needed to enhance the individual components using precisely selected single 
pulses to make the relevant components more prominent. We have employed three 
different techniques (mentioned below) to generate the most prominent 
realization of each component in the pulsar profile. The components once 
identified were classified into core and the conal types and their respective 
widths at each frequency were measured at the 50\% level of the peak intensity 
($W_{core}$ and $W_{cone}$). The 50\% level was selected as a representative 
width of the component since any lower level (like 25\% or 10\%) are usually 
contaminated by the adjacent components making the estimation difficult in a 
large number of components.

\subsection{\large \bf Identifying Profile Components} \label{sec:tech_1}
\noindent
Here we discuss the three different techniques that were used to measure 
component widths using the single pulse analysis. Before the single pulses were
averaged the baseline level from each single pulse was removed and the 
integrated profile peaks were normalized to unity after the components were 
formed. 

\begin{figure*}
\begin{center}
\begin{tabular}{@{}lr@{}}

{\mbox{\includegraphics[angle=0,scale=0.25]{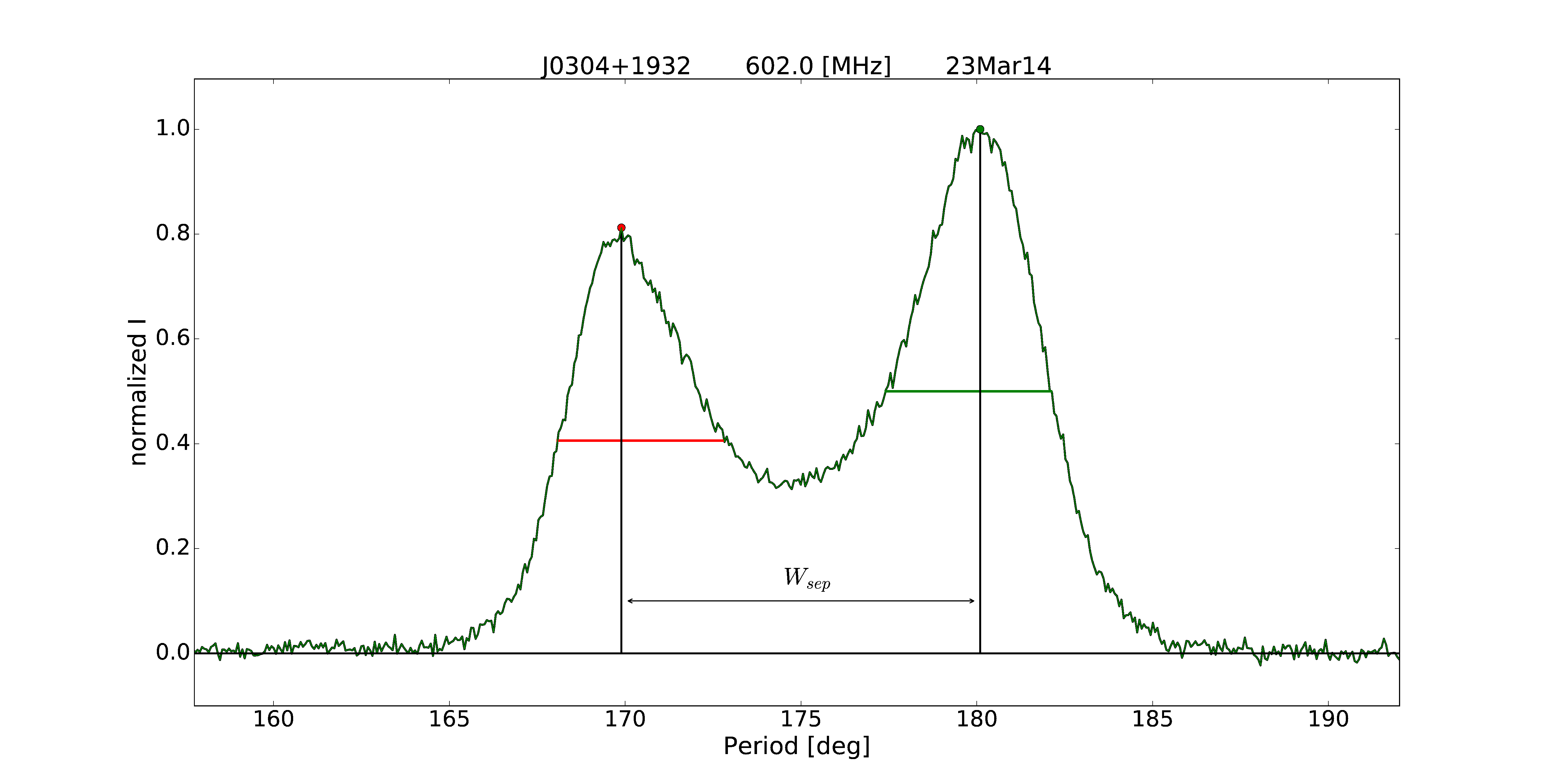}}}\\
\end{tabular}
\end{center}
\caption{The figure shows the measured width ($W_{cone}$) for the conal 
components and the separation ($W_{sep}$) between the peaks in the pulsar 
J0304+1932. The green and red points on the graph correspond to the peaks of 
the individual components, while the horizontal lines of the same color 
correspond to the measured component widths at 50\% level of the peak. The 
widths were estimated using the integrated profile (see section 
\ref{sec:option_w}). The black vertical lines indicate the measured separation 
between the peaks.
\label{full}}
\end{figure*}

1. {\bf Integrated Profile}\\
\noindent
In this method the average profiles were produced with an additional 
modification of selecting only significant single pulses with peaks above five 
times the rms noise level of the off-pulse baseline. This enhanced the signal 
to noise ratio (SNR) of the components especially in pulsars which showed 
nulling \citep{bas17}. This technique was most widely used in this work with 89
pulsars at 333~MHz and 112 pulsars at 618~MHz where the components were 
characterized. Figure~\ref{full} illustrates the integrated profile for the 
pulsar J0304+1932 where we used this method.\\

\begin{figure*}
\begin{center}
\begin{tabular}{@{}lr@{}}
{\mbox{\includegraphics[angle=0,scale=0.25]{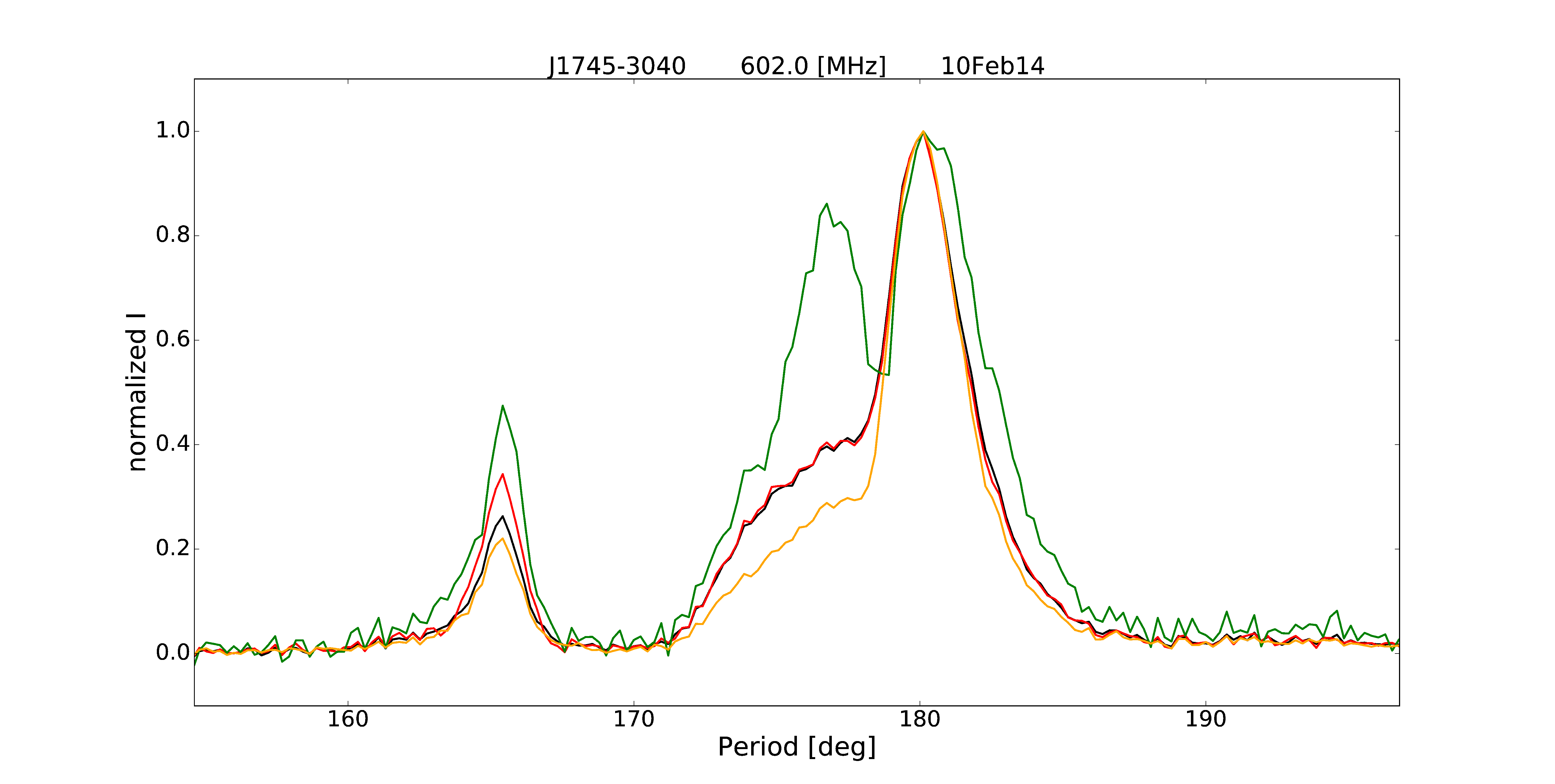}}}\\
\end{tabular}
\end{center}
\caption{The figure shows the component estimation in the pulsar J1745$-$3040. 
The black curve shows the integrated profile. The yellow, green and red curves 
correspond to the profiles used to measure the first, second and third 
component using the second method described in section \ref{sec:tech_1}. The 
second component could be clearly distinguished only after using these 
specialized techniques. The leading component is made up of sporadic emission 
and is unlikely to be a fully formed conal component. The width of the 
component is smaller than the typical conal width and marked separately as a
yellow dot in figure~\ref{result_618}.
\label{m2}}
\end{figure*}

\noindent
2. {\bf Averaging Subpulses}\\
\noindent
In this method we separated out the components in 2 pulsars at 333~MHz and 3 
pulsars at 618~MHz which appear merged in the average profile. The subpulses 
corresponding to the components were seen to separate out in the single pulses 
and with jittering in phase. The peaks of each component were identified in the
average profile using a peak detection technique\footnote{The peaks were 
estimated using the minima of second derivatives of the profile curve.}. 
Template windows of width 2.45\degr$P^{-0.5}$ and center phase corresponding to
the component peaks were set up. Any single pulse exceeding three times the 
off-pulse noise levels in each of these windows were considered for the 
respective average components. In addition a further criterion for averaging 
was that the peak of the subpulse within the window was close to the central 
phase (within 5-10 longitude bins). Finally, all relevant single pulses were 
averaged to generate a profile with the relevant component prominently 
detected. A separate profile for each individual component was generated in 
this technique. In figure \ref{m2} the results of this exercise for the pulsar 
J1745$-$3040 is shown, where the second component becomes prominent after 
applying this technique. The leading component of this pulsar do not form a
fully formed conal component, resembling a pre-cursor \citep{bas15}, and has 
not been used for component width analysis.\\

\begin{figure*}
\begin{center}
\begin{tabular}{@{}lr@{}}
{\mbox{\includegraphics[angle=0,scale=0.25]{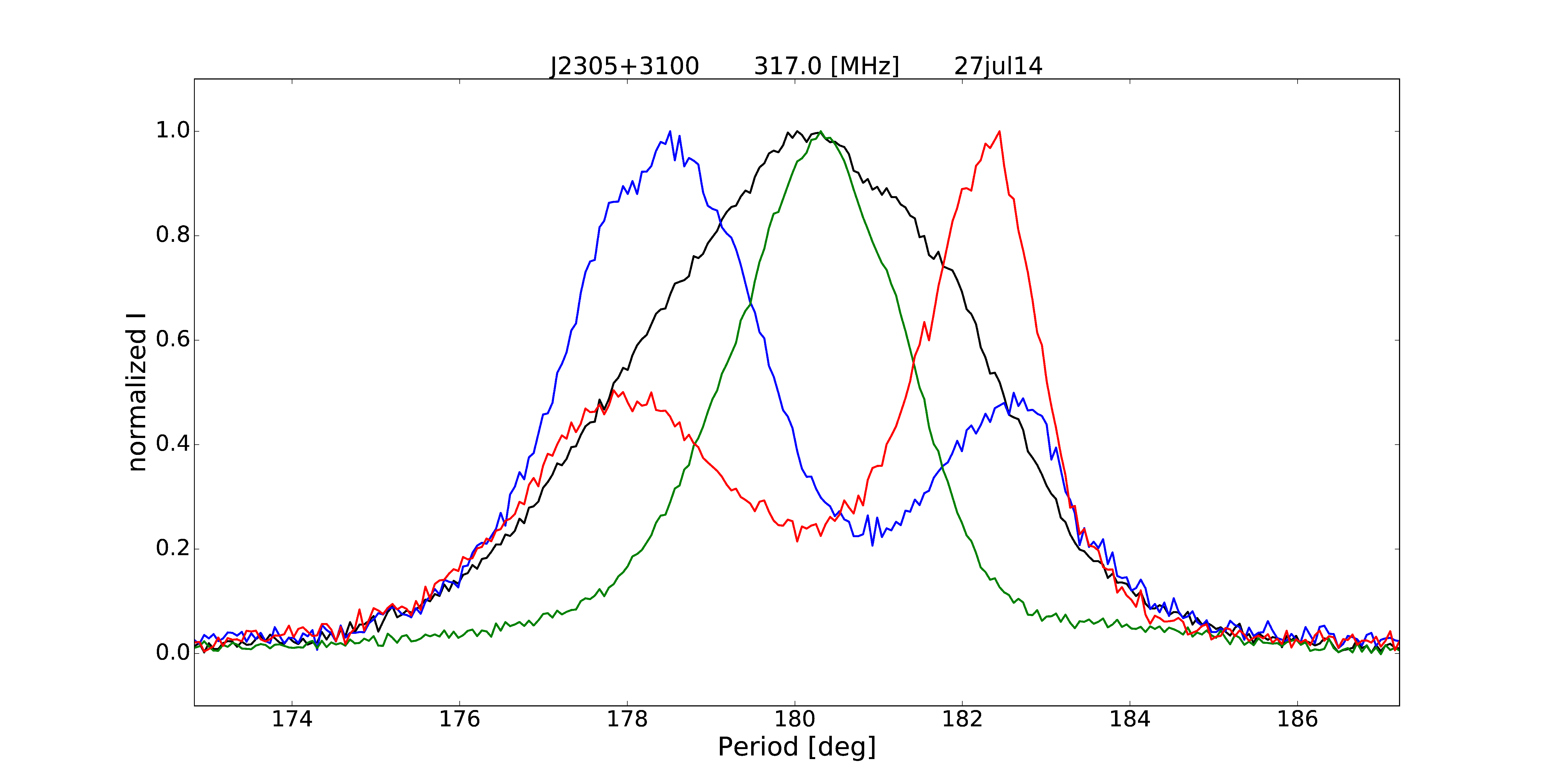}}}\\
\end{tabular}
\end{center}
\caption{This figure demonstrates the use of the third method described in 
section \ref{sec:tech_1} for the pulsar J2305+3100. This method was used for 
pulsars showing subpulse drifting with prominent drift bands. The components 
were well separated at the single pulse level but shifted across the window 
giving a smeared average profile (black line). As shown by the red, blue, green
profiles, our technique was successful in separating the components and 
estimating their widths as well as the separation between them.
\label{m3}}
\end{figure*}

\noindent
3. {\bf Averaging Peaks in Window}\\ 
\noindent
This technique was used to estimate the components in 20 pulsars at 333~MHz and 
19 pulsars at 618~MHz. This was mainly useful for subpulses which showed 
prominent drift bands and systematic shift in phase within the pulse window. In
order to measure the component widths a technique was devised to average the 
peaks appearing at similar phases. Inside the profile windows, 5-10 bins wide, 
were set up and were spread contiguously across the whole profile. All 
significant subpulses (peak intensity greater than five times the baseline rms)
with peaks within the relevant window were averaged to form a profile 
corresponding to that window. All such profiles with at least $10\%$ of the 
total pulses were used for subsequent analysis. The profile separated out into 
individual components as shown in  figure~\ref{m3} for the pulsar J2305+3100.
By this method we generated several representative profiles which had one or 
components. We measure the widths of all these components and produced an 
average component width for the pulsar. Similarly, we estimated the separation 
between components for all relevant cases and estimated an average $W_{sep}$.\\
 
\subsection{\large \bf Measuring Component Widths}\label{sec:option_w}
\noindent
Using the analysis scheme discussed above we were able to develop the best 
possible profile for the individual components. However, it was not always 
possible to clearly separate out the components in all cases. But in many 
of these cases it was still possible to find an estimate of the widths using 
certain fitting procedures. We employed three different techniques for 
measuring the 50\% widths of the components which we describe below.\\

\noindent
1. {\bf Full Width}\\
\noindent
This was the commonly used technique in our analysis where the 50\% of peak
intensity on either side of the component could be clearly measured (see 
figure~\ref{full}). The estimated width was the separation between the 50\% 
points on either side of the peak. A total of 99 component widths at 333~MHz 
and 123 widths at 618~MHz were measured using this technique.\\ 


\noindent
2. {\bf Half Width}\\
\noindent
This method was used to estimate the widths of 76 components at 333~MHz and 82 
components at 618~MHz. This was utilized when the peaks were clearly seen but 
one side was not well separated from the adjacent component. The separation of the 
peak from the 50\% point of the well resolved side was calculated and 
multiplied by two, to obtain an approximate estimate of the component width.\\

\noindent
3. {\bf Gaussian Fits} \\
\noindent
In a small number of cases, 11 components at 333~MHz and 20 components at 
618~MHz, the peak was not clearly seen. In such cases a Gaussian function was 
fitted to the component and the full width at half maximum (FWHM) of the 
Gaussian was used as an estimate of the width. However, the Gaussian fit is not
always an accurate model for the subpulses and the widths in such cases should 
be used with caution.

\subsection{\large \bf Separation Between Adjacent Components}\label{sec:sep}
\noindent
The separation between adjacent components ($W_{sep}$) in profiles with more 
than one component was calculated. We used the peak positions estimated in the 
previous subsection for each component to find the separation between them 
(see figure \ref{full} for a schematic). In some cases the exact location of 
the peak was uncertain due to lower sensitivity of the emission and a Gaussian 
fit was used to approximate the peak position. 

\subsection{\large \bf Error Estimation} \label{sec:err}
\noindent
The error ($\sigma$) in estimating the component width was evaluated as 
\citep{kij03}: 
\begin{equation} \label{eq6}
\sigma=res\times \sqrt{1+\left(\frac{rms}{I}\right)^2},
\end{equation}
where $res$ is the longitude resolution in degrees, $rms$ corresponds to the 
baseline noise levels and $I$ is the measured signal, in our case $I$ 
corresponds to peak intensity. In case of the separation between components, 
the error in estimating each peak position was determined and the overall error
was calculated by propagating the individual errors.

\section{\large Results} \label{sec:res}
\noindent
In Table~\ref{tabela} we have presented the results of our analysis for the 123
pulsars observed at two frequencies 333~MHz and 618~MHz. We have determined the
number of profile components and its classification at each frequency and 
measured their widths wherever possible. In addition we have also determined 
the separation between adjacent components. Additionally, the table also lists 
the measurement schemes employed for every component (see section 
\ref{sec:tech_1} and \ref{sec:option_w} for details). 

\subsection{\large Component Widths and Separation} 
\begin{figure*}
\begin{center}
\begin{tabular}{@{}lr@{}}
{\mbox{\includegraphics[angle=0,scale=0.3]{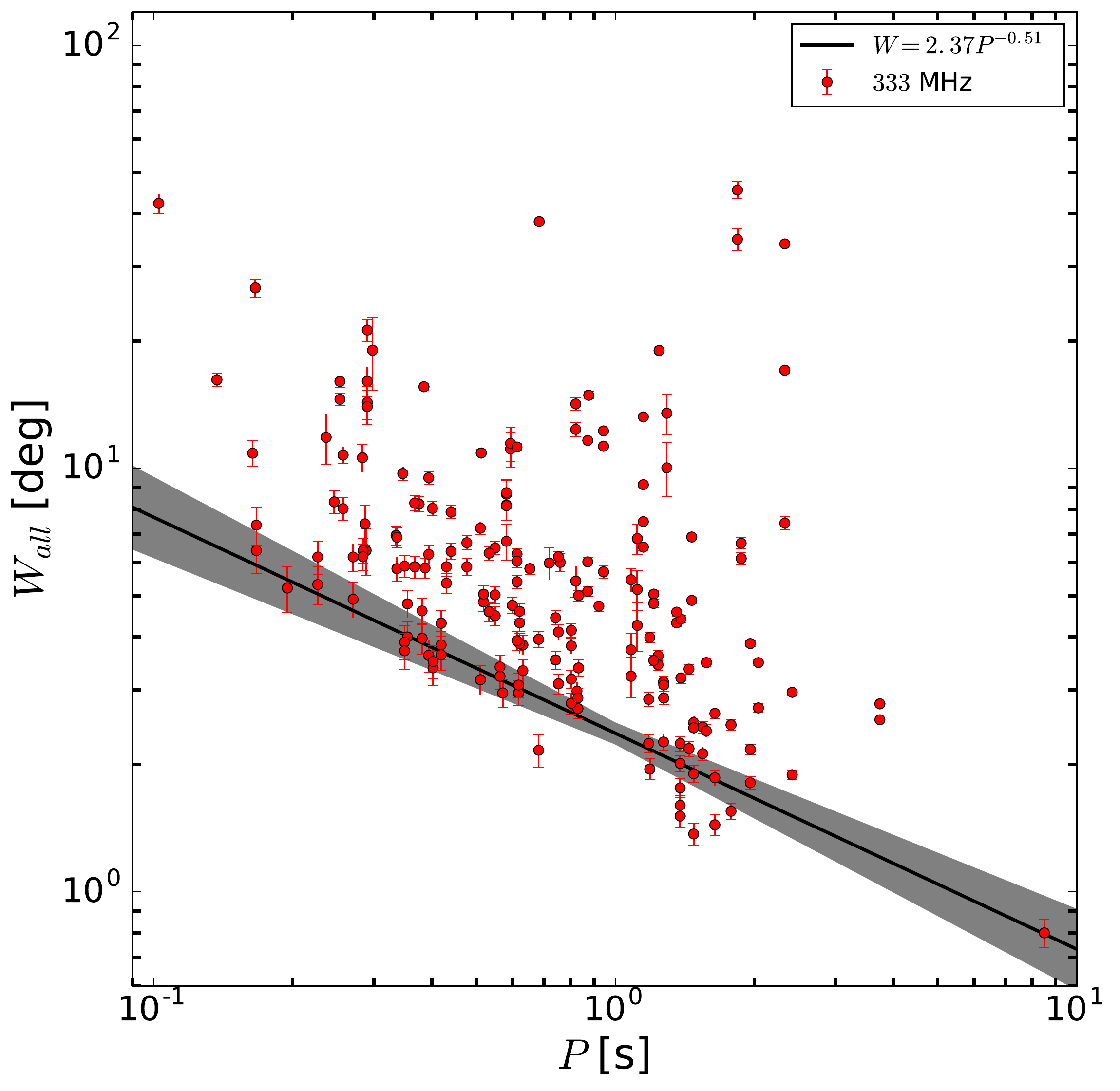}}}&
{\mbox{\includegraphics[angle=0,scale=0.3]{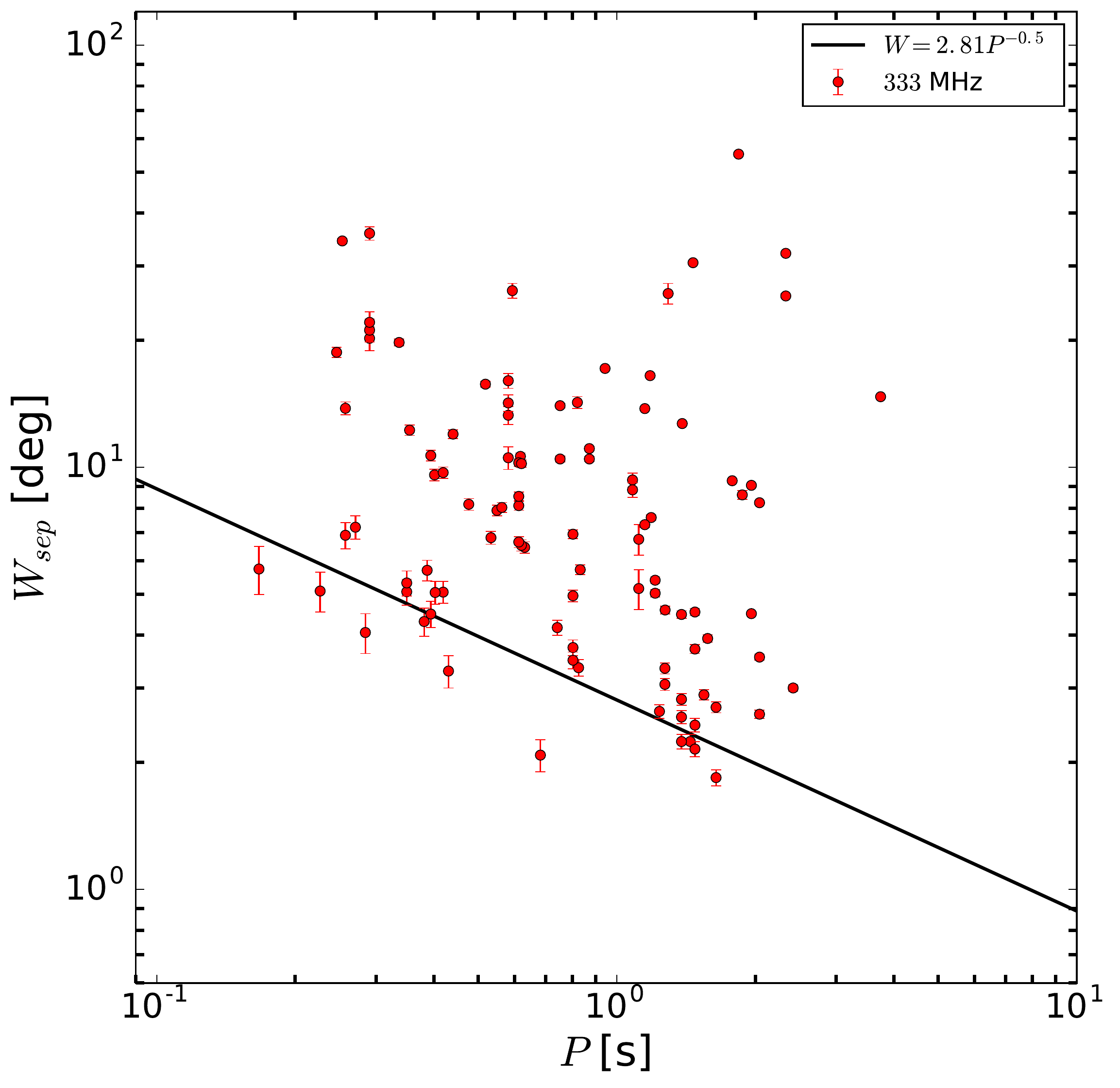}}}\\
{\mbox{\includegraphics[angle=0,scale=0.3]{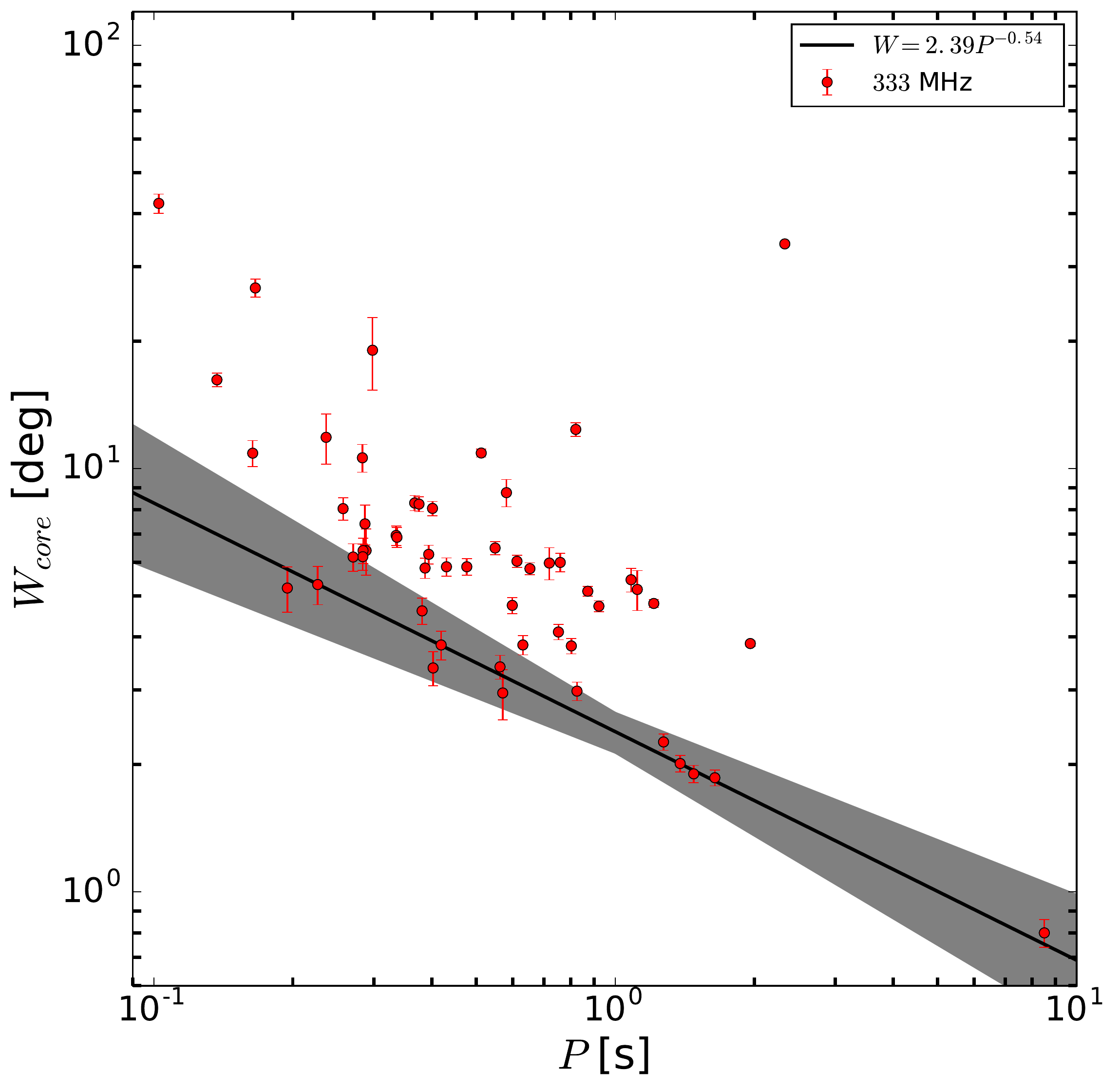}}}&
{\mbox{\includegraphics[angle=0,scale=0.3]{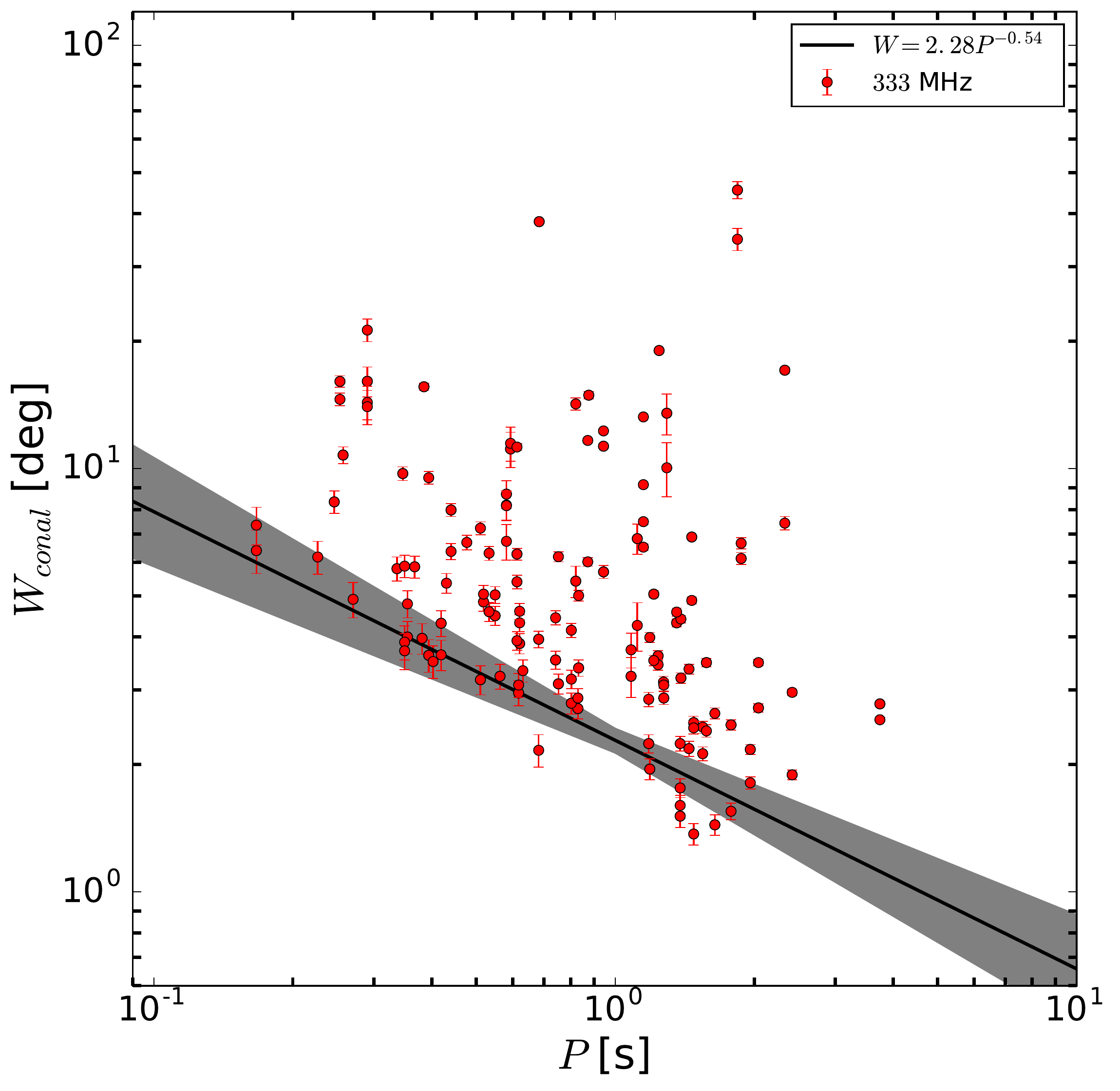}}}\\
\end{tabular}
\end{center}
\caption{In this figure we plot the widths $W_{all}$, $W_{core}$ and 
$W_{cone}$ as well as the separation $W_{sep}$ as a function of the pulsar 
period ($P$) at 333 MHz. On the upper left panel both core and conal components
are shown together whereas on the lower left panel the core component is 
plotted separately and likewise for the conal components on the lower right 
panel. On the upper right panel $W_{sep}$ is plotted against $P$. The black 
line in each plot is our estimate of the lower boundary line using quantile 
regression (see section~\ref{sec:bound}) with the grey area representing the 
errors from our fits. In the case of $W_{sep}$ we were not able to constrain 
the lower boundary line using the fitting process and the boundary was obtained
by assuming a $P^{-0.5}$ dependence.}
\label{result_333}
\end{figure*}

\begin{figure*}
\begin{center}
\begin{tabular}{@{}lr@{}}
{\mbox{\includegraphics[angle=0,scale=0.3]{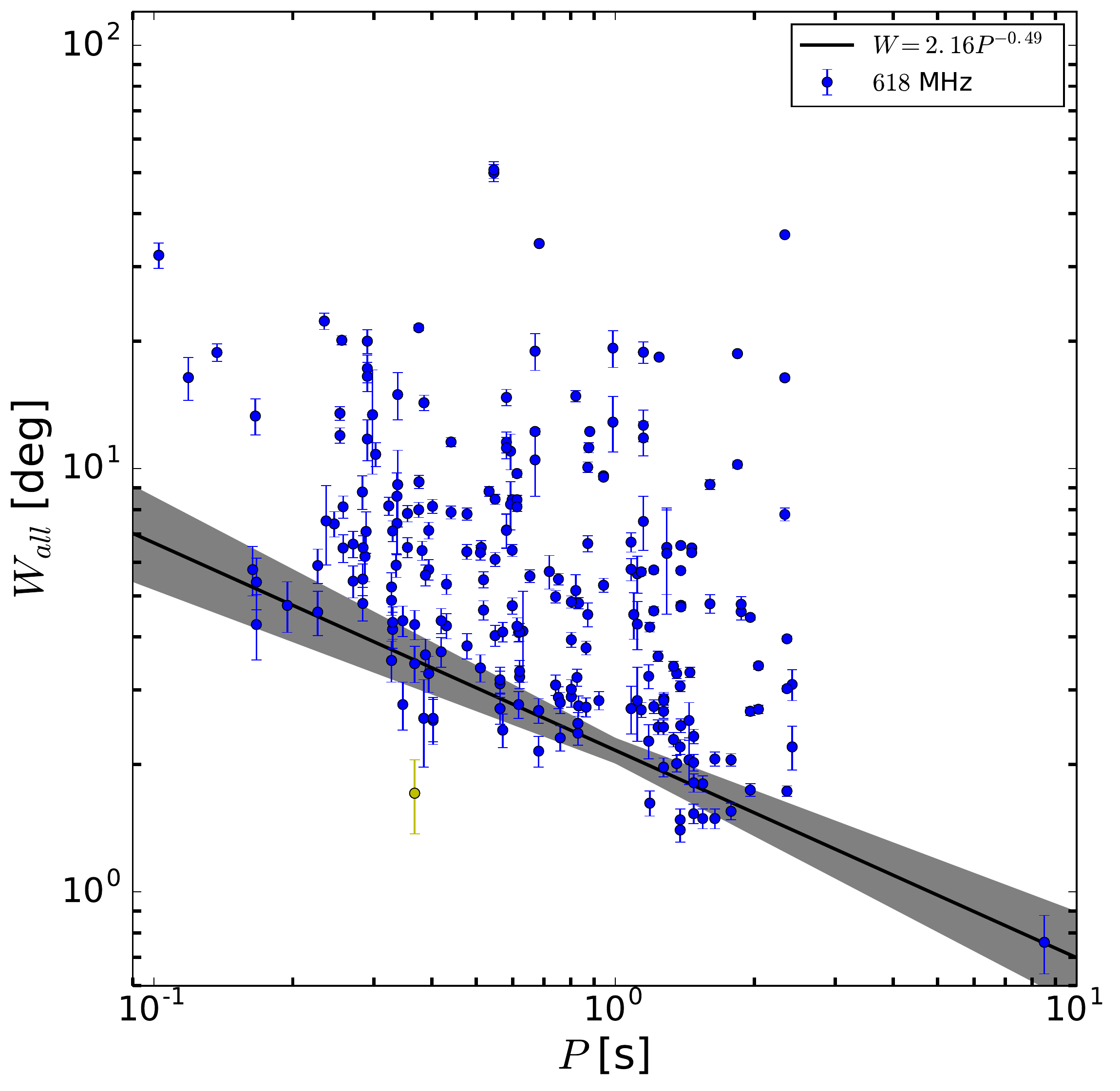}}}&
{\mbox{\includegraphics[angle=0,scale=0.3]{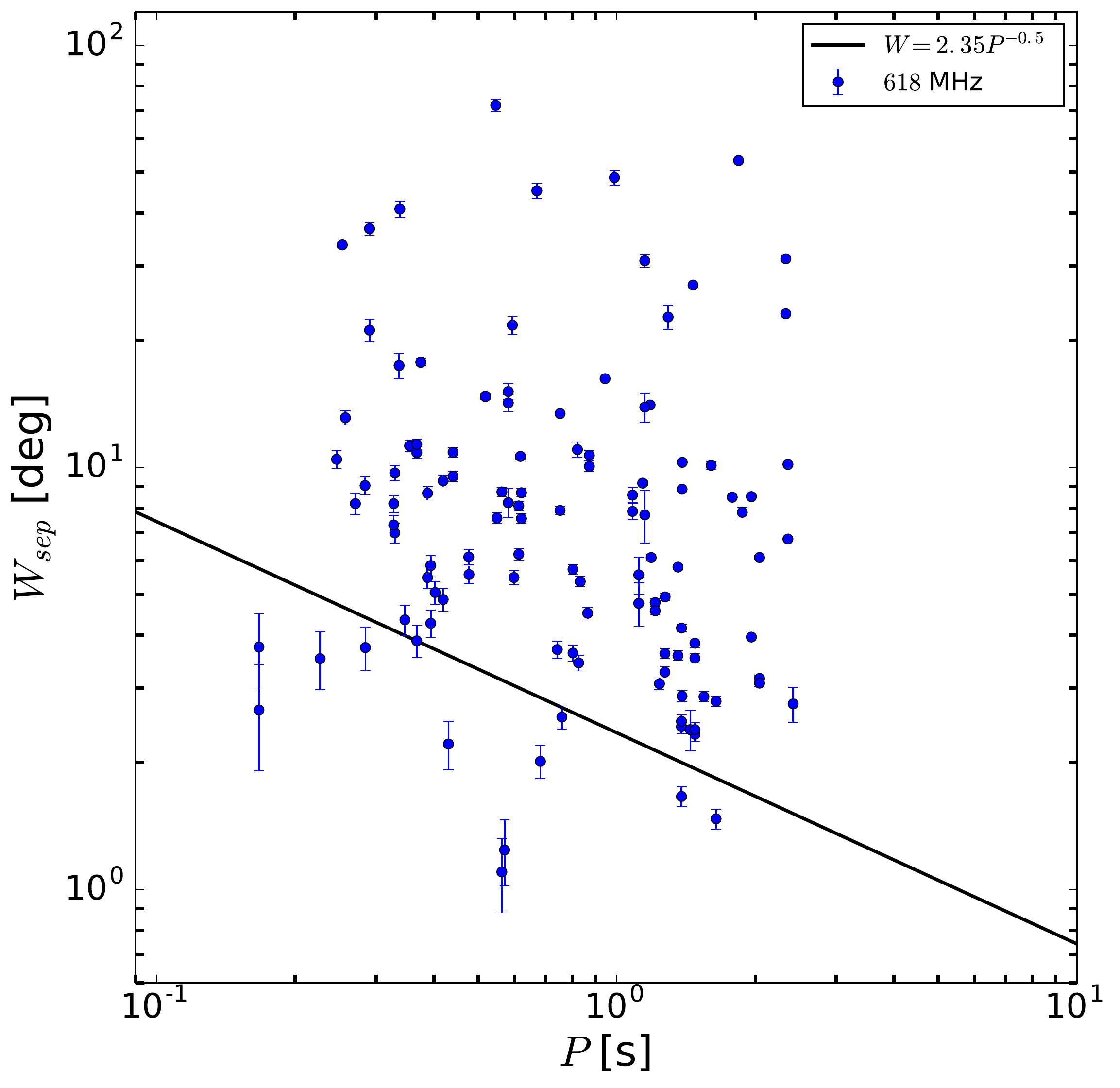}}}\\
{\mbox{\includegraphics[angle=0,scale=0.3]{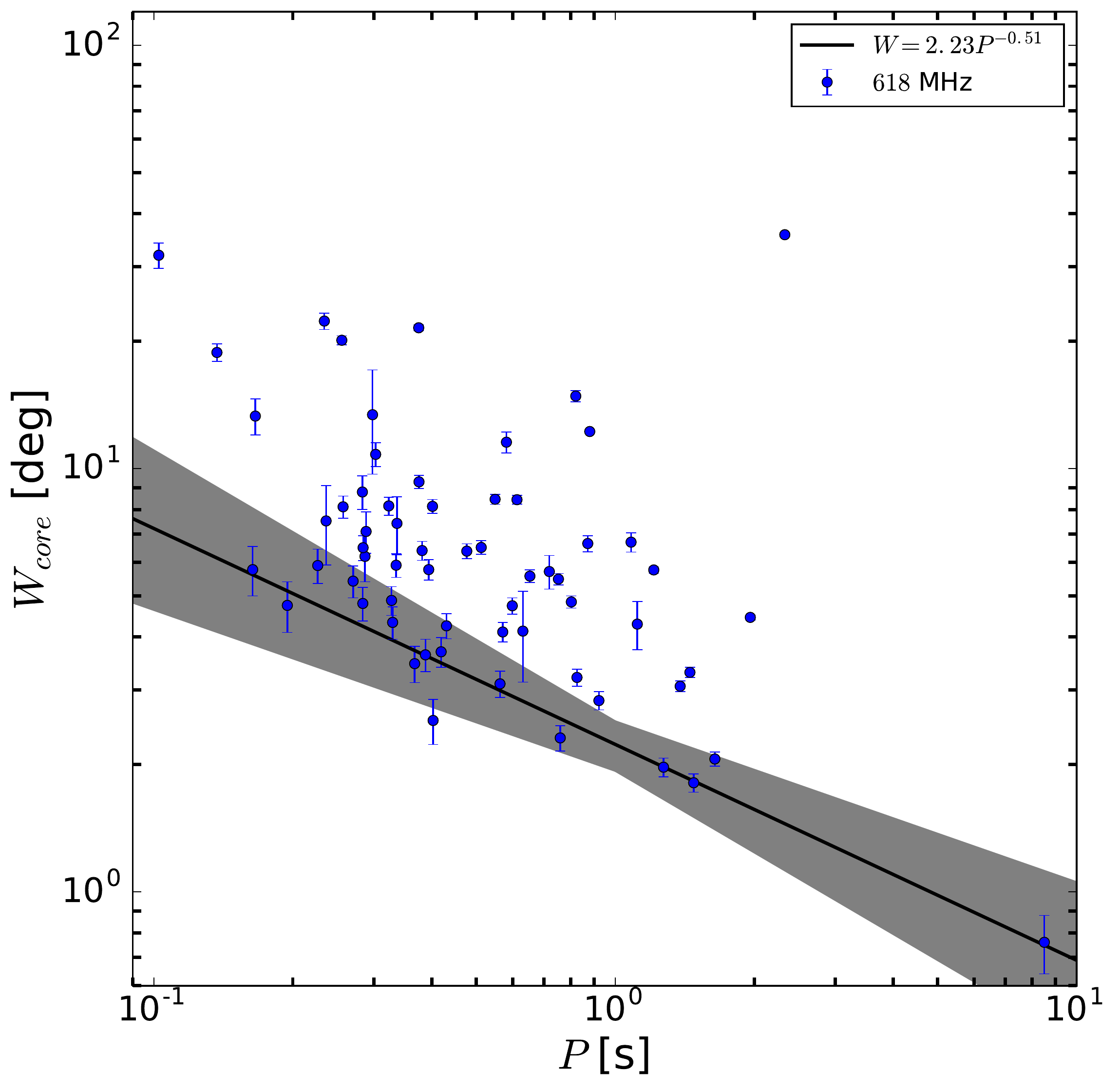}}}&
{\mbox{\includegraphics[angle=0,scale=0.3]{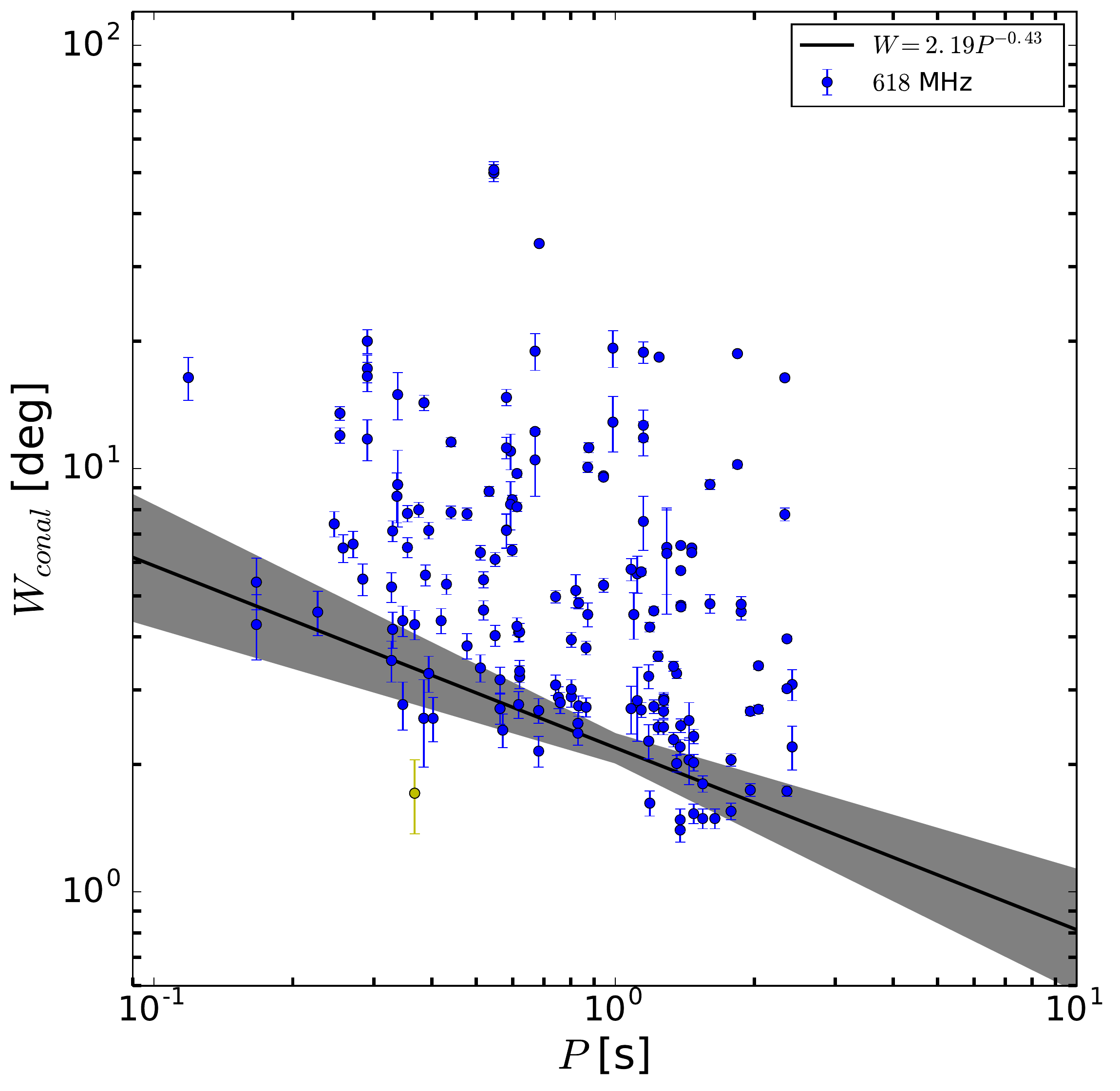}}}\\
\end{tabular}
\end{center}
\caption{In this figure we plot the widths $W_{all}$, $W_{core}$ and 
$W_{cone}$ as well as the separation $W_{sep}$ as a function of the pulsar 
period ($P$) at 618 MHz. The yellow point corresponds to the leading component 
in PSR J1745$-$3040 which is not a fully formed conal component and has a lower
width. On the upper left panel both core and conal components are shown 
together whereas on the lower left panel the core component is plotted 
separately and likewise for the conal components on the lower right panel. On 
the upper right panel $W_{sep}$ is plotted against $P$. The black line in each 
plot is our estimate of the lower boundary line using quantile regression (see 
section~\ref{sec:bound}) with the grey area representing the errors from our 
fits. In the case of $W_{sep}$ we were not able to constrain the lower boundary
line using the fitting process and the boundary was obtained by assuming a 
$P^{-0.5}$ dependence.}
\label{result_618}
\end{figure*}

\begin{figure*}
\begin{center}
\begin{tabular}{@{}lr@{}}
{\mbox{\includegraphics[angle=0,scale=0.3]{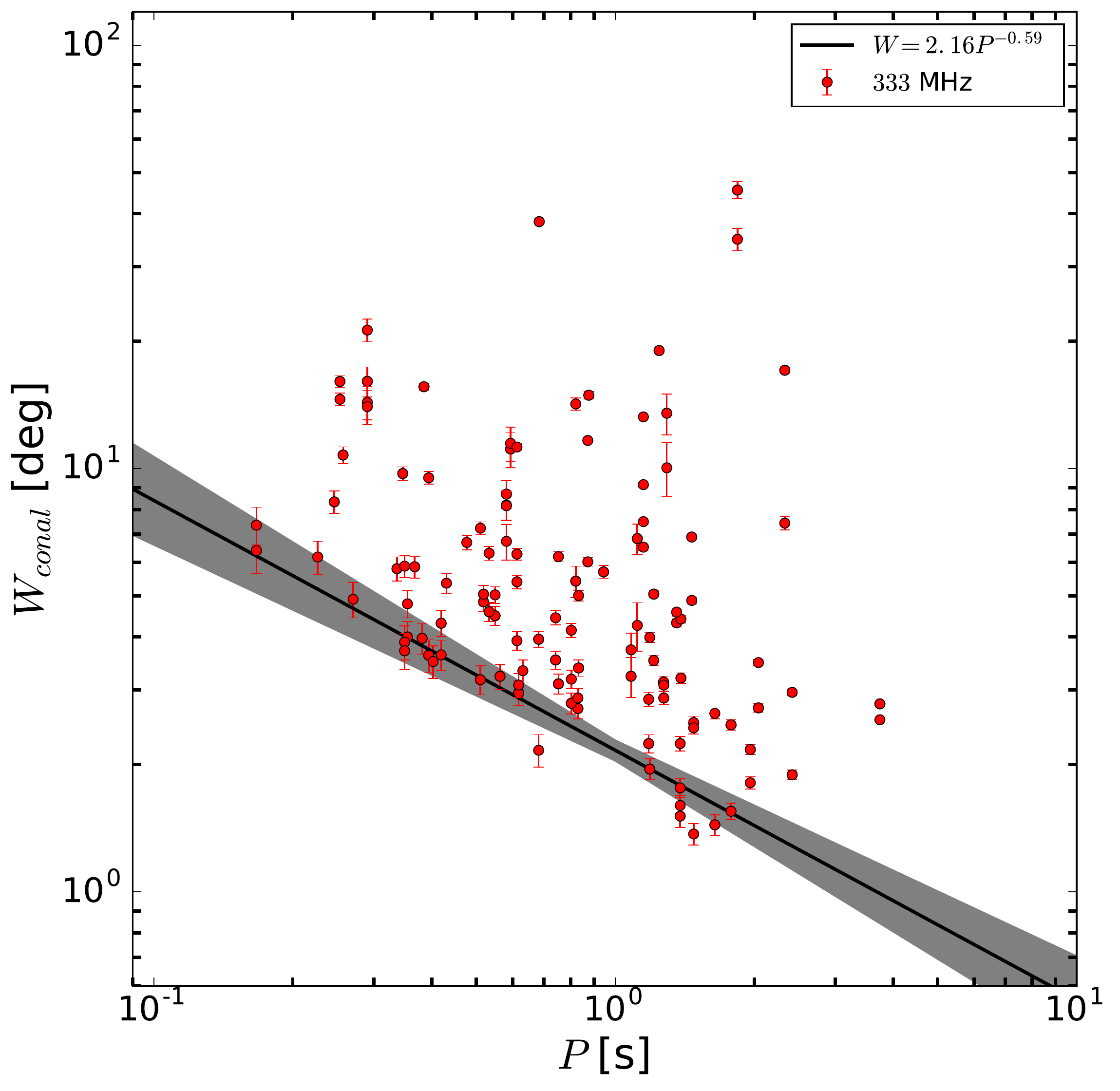}}}&
{\mbox{\includegraphics[angle=0,scale=0.3]{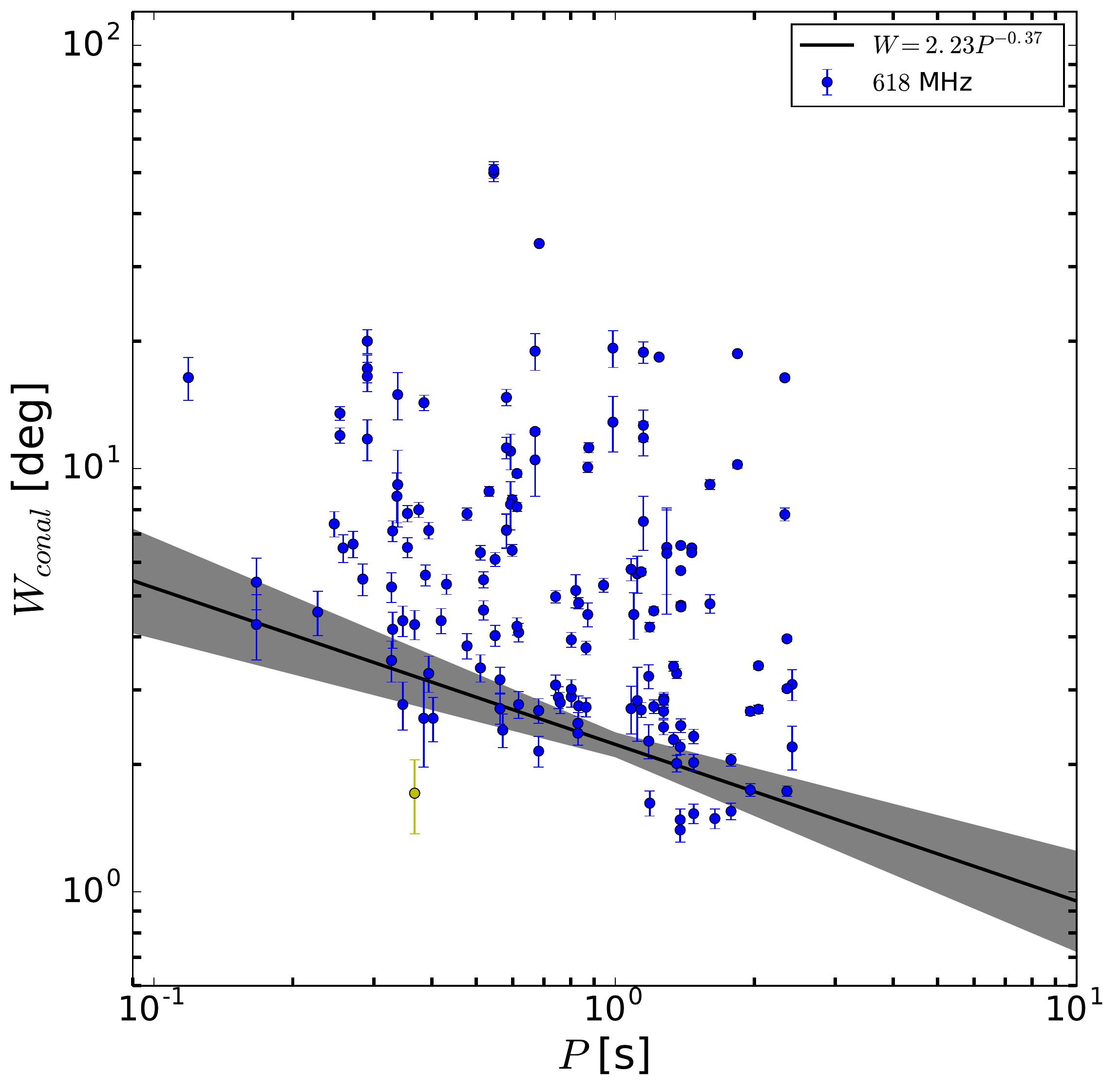}}}\\
{\mbox{\includegraphics[angle=0,scale=0.3]{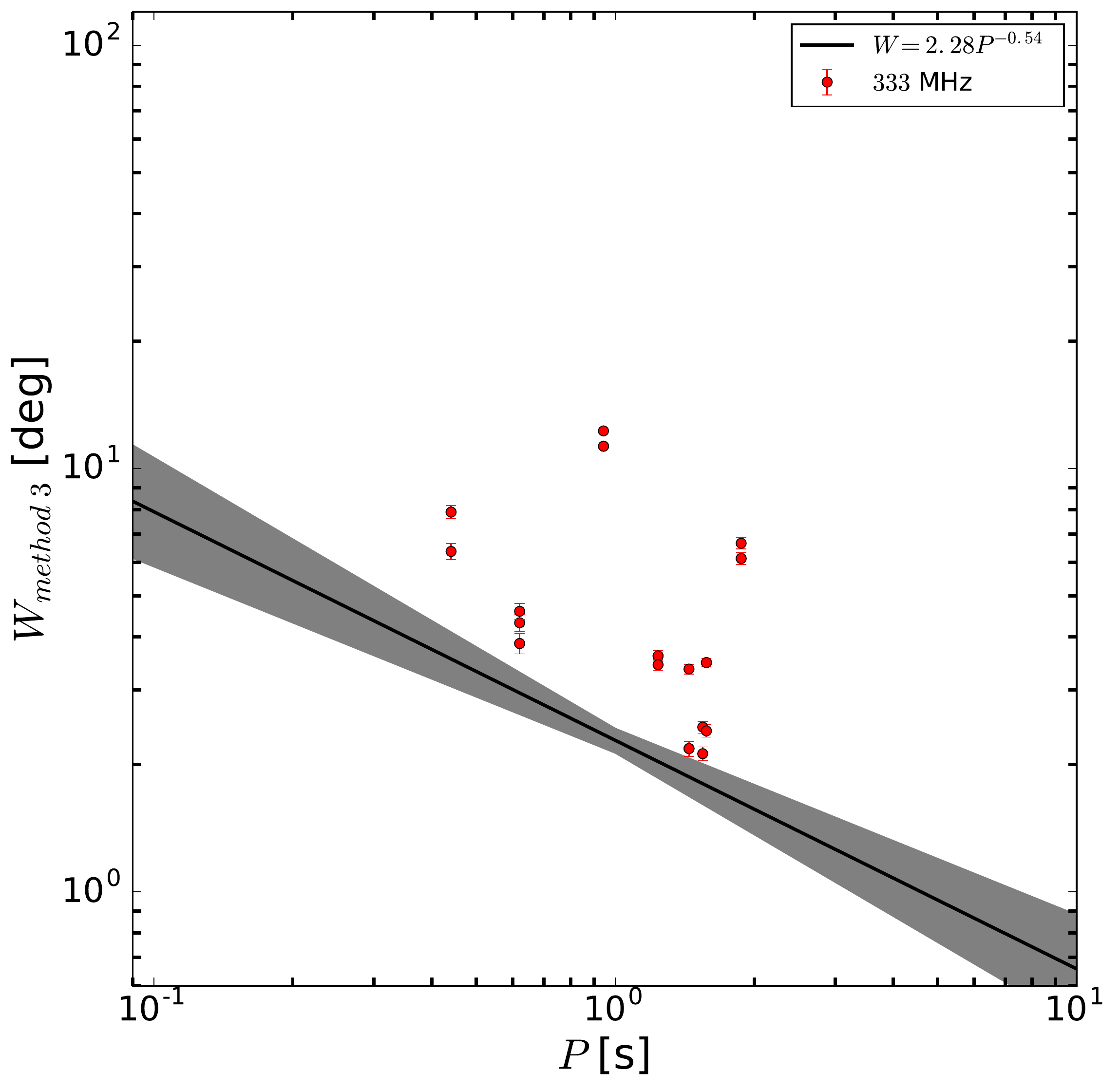}}}&
{\mbox{\includegraphics[angle=0,scale=0.3]{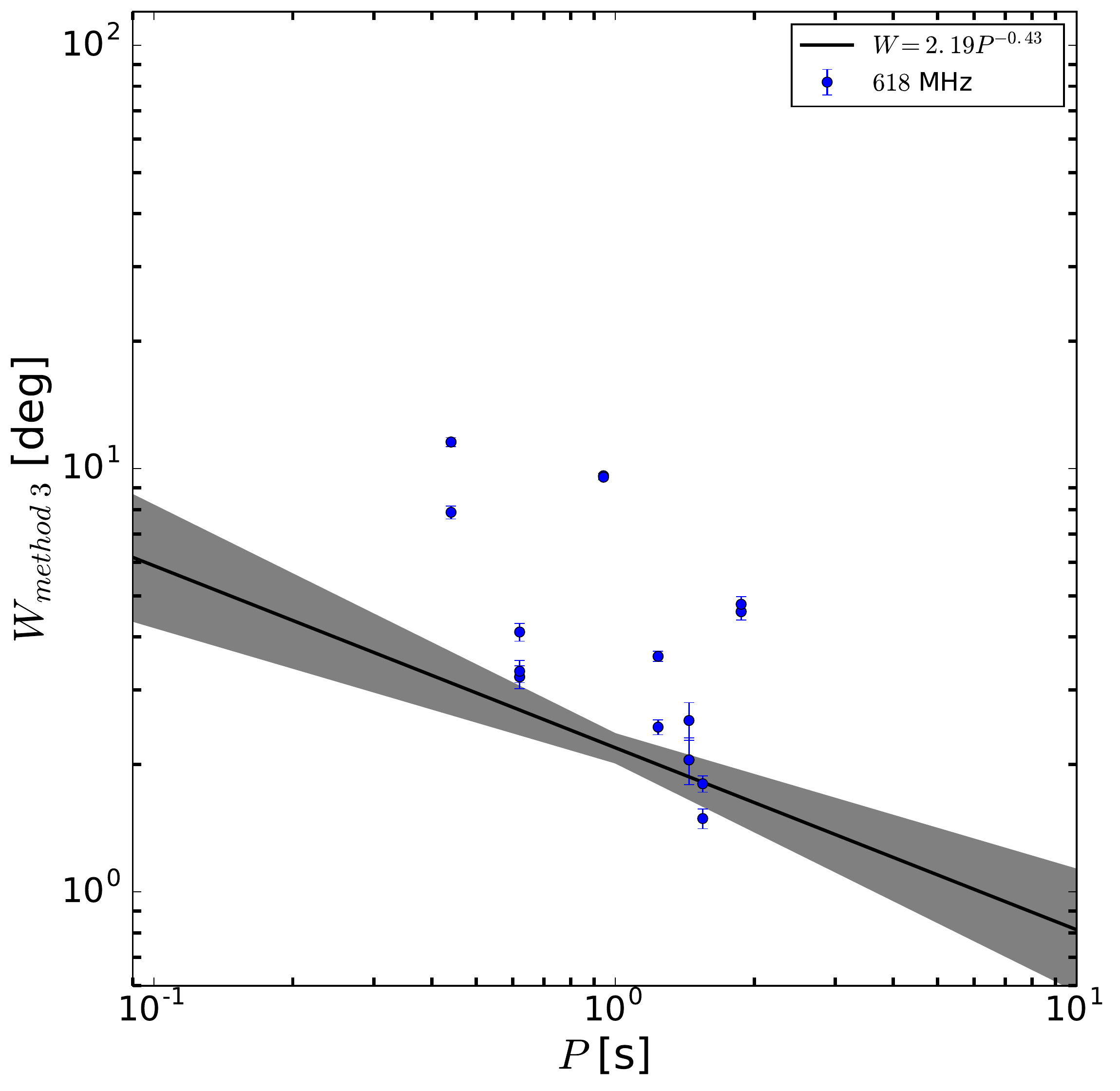}}}\\
\end{tabular}
\end{center}
\caption{In this figure we separately show that the values of the 
conal widths for pulsars showing subpulse drifting, measured using method 3~in 
section \ref{sec:tech_1}, do not dominate the boundary line estimates. The top 
figures show the conal width distributions without the drifting widths at 333 
MHz (top left panel) and 618 MHz (top right panel) respectively. The boundary 
line is estimated in each case using the method of quantile regression and is 
identical to the overall estimates within measurement errors. The bottom panels
show the drifting widths at 333 MHz (bottom left panel) and 618 MHz (bottom 
right panel) respectively. The lower boundary could not be estimated 
statistically for just the drifting cases due to the small number of 
measurements. However, the lower boundary from figures \ref{result_333} and 
\ref{result_618} is shown in these figures. It is seen that the estimated widths
do not occupy the boundary line but are widely distributed.}
\label{result_drift}
\end{figure*}

\noindent
The widths were measured for 418 components in total, including 191 widths at 
333~MHz and 227 widths at 618~MHz. There were 117 $W_{core}$ in total out of 
which 54 $W^{333{\rm MHz}}_{core}$ and 63 $W^{618{\rm MHz}}_{core}$. 
Correspondingly, we measured 301 $W_{cone}$ out of which 137 
$W^{333{\rm MHz}}_{cone}$ and 164 $W^{333{\rm MHz}}_{cone}$. In addition 
$W_{sep}$ had 217 measurements, with 100 $W^{333{\rm MHz}}_{sep}$ and 117 
$W^{618{\rm MHz}}_{sep}$. The period dependence of $W_{core}$, $W_{cone}$ and 
$W_{sep}$ in logarithmic scales for each frequency are shown in 
figure~\ref{result_333},\ref{result_618}. There are four separate plots 
corresponding to all the measured widths, $W_{all}$ (top left panel), all 
$W_{sep}$ (top right panel), $W_{core}$ (bottom left panel) and $W_{cone}$ 
(bottom right panel). In each case it is clear that a lower boundary exists for
the widths as well as the separations, though the boundary is less tightly 
constrained for $W_{sep}$. In addition figure~\ref{result_drift} 
separately shows the conal widths estimated for the subpulse drifting pulsars 
using method 3~in section~\ref{sec:tech_1}. It is seen that the boundary 
estimates are not affected by these widths. A statistical approach using 
quantile regression (see appendix~\ref{sec:quant} for a discussion on quantile 
regression) was employed to determine the LBL in each case.

\subsection{\large The Lower Boundary Line}\label{sec:bound}
    
\begin{table*}
\caption{THE LOWER BOUNDARY LINE: $W$ = $W_B$\degr$P^{-b}$  \label{tab_fit}}
\begin{center}

\begin{tabular}{c | c | cc | cc | cc}
\tableline
 \multicolumn{2}{c}{}  & \multicolumn{2}{c}{333 MHz} & \multicolumn{2}{c}{618 MHz} & \multicolumn{2}{c}{Both Freq.} \\
\tableline
  &  & $W_B$ & $b$ & $W_B$ & $b$ & $W_B$ & $b$  \\
\tableline
                 & $W_{core}$ & 2.39$\pm$0.26 & 0.54$\pm$0.11 & 2.23$\pm$0.30 & 0.51$\pm$0.13 & 2.28$\pm$0.19 & 0.51$\pm$0.08 \\ 
                 &        &               &               &               &               &               &               \\ 
 Comp. Width & $W_{cone}$ & 2.28$\pm$0.15 & 0.54$\pm$0.10 & 2.19$\pm$0.17 & 0.43$\pm$0.11 & 2.24$\pm$0.11 & 0.49$\pm$0.07 \\ 
                 &        &               &               &               &               &               &               \\ 
                 & $W_{all}$ & 2.37$\pm$0.13 & 0.51$\pm$0.07 & 2.16$\pm$0.14 & 0.49$\pm$0.08 & 2.25$\pm$0.09 & 0.51$\pm$0.05 \\ 
\tableline
                 &        &               &               &               &               &               &               \\
    $W_{sep}$    &        &      2.81     &      0.5      &      2.35     &      0.5      &      2.66     &      0.5      \\ 
                 &        &               &               &               &               &               &               \\ 
\tableline
                 &        &               &               &               &               &               &               \\
  Prof. Width &        &      3.30     &      0.5      &      2.97     &      0.5      &      2.99     &      0.5      \\ 
                 &        &               &               &               &               &               &               \\ 
\tableline
\end{tabular}
\end{center}
\end{table*} 

\noindent
The LBL is of the form $W_{B} P^{-b}$, where $W_{B}$ is the boundary width and 
$b$ the period dependence. Table~\ref{tab_fit} shows the estimates of the LBL 
for each category of measured widths. The first row represents the estimates of
LBL for $W_{core}$, $W_{cone}$ and $W_{all}$ at each frequency using quantile 
regression. The $b$ value in each case is close to $0.5$ suggesting that 
the $P^{-0.5}$ dependence is consistent for the component widths. We use the 
period dependence to exhibit the $P^{-0.5}$ relation in subsequent discussions 
as a close approximation to the measured dependence. The boundary value $W_{B}$
was lower than 2.45\degr~which was the previously estimated boundary for the 
core widths at 1~GHz. The frequency evolution of component widths suggests that
the measurements at 333 MHz and 618 MHz should be greater than the 
corresponding values at 1~GHz. However, the measured $W_B$ at 333 MHz was 
greater than at 618 MHz in accordance with our expectations. Assuming a 
frequency dependence of component width ($\propto \nu^{\gamma}$), the estimated
$\gamma$ is -0.15$\pm$0.01 and the boundary at 1~GHz is $W_{B}^{1 {\rm GHz}}$ =
2.01$\pm$0.15\degr~which is smaller than the previous estimates. In addition, 
the results also show that $W_{B,core}$ (2.28$\pm$0.19\degr) and $W_{B,cone}$ 
(2.24$\pm$0.11\degr) are similar. There was a possibility that the lower 
boundary line was affected by the drifting widths measured using method 3~in 
section \ref{sec:tech_1}. To eliminate this possibility the boundary lines were 
estimated separately by removing the drifting widths. The estimated conal
boundary line for the entire population without the drifting widths are 
$W_{B,cone,nd}^{333 {\rm MHz}}$ = 2.16$\pm$0.12\degr and 
$W_{B,cone,nd}^{618 {\rm MHz}}$ = 2.23$\pm$0.14\degr, respectively. These 
values are similar to the overall boundaries (within measurement errors) and 
show that the widths of the drifting components do not bias the boundary line. 

The statistical method was unable to constrain the period dependence in 
$W_{sep}$. We nonetheless estimated a boundary for these quantities as well by 
assuming a $P^{-0.5}$ dependence and estimating boundary below which only 10\%
points were present (the 0.1 quantile level, see appendix). The estimated 
boundaries of $W_{sep}$ at 333~MHz, 618~MHz and all combined are shown in 
second row of Table~\ref{tab_fit}. Additionally, the lower boundary for 
$W_{prof}$ in the average profiles \citep{mit16b} was also estimated at the 
0.1 quantile level (assuming a $P^{-0.5}$ dependence) and are shown in third 
row of Table~\ref{tab_fit}.

\section{\large Discussion}\label{sec:dis}
\subsection{\large Period dependence of Component widths} \label{sec:modl}
\begin{figure*}
\begin{center}
{\mbox{\includegraphics[angle=0,scale=0.3]{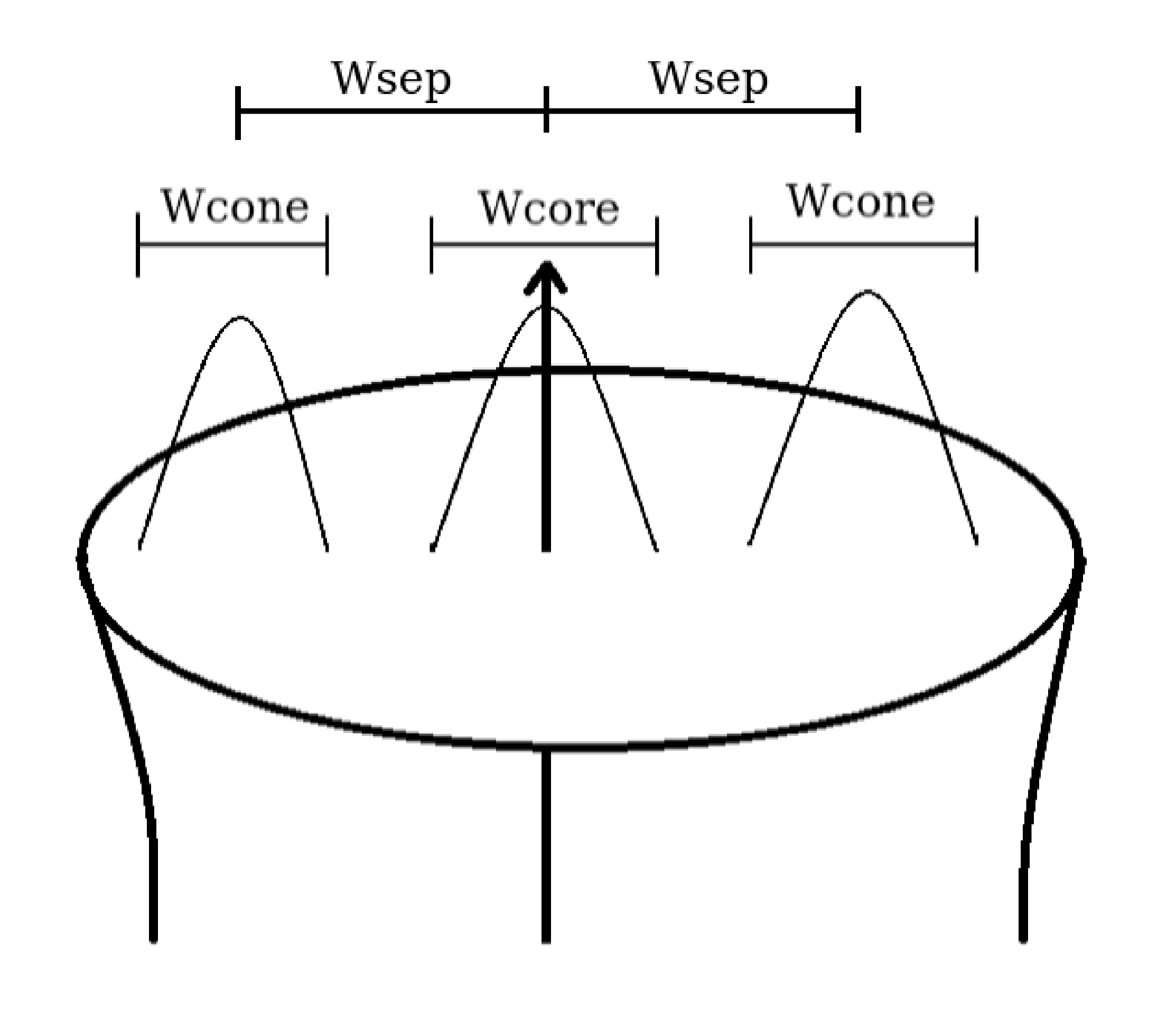}}}
\end{center}
\caption{\small The figure shows a schematic of the core and conal components 
along the dipolar open field line region in a normal pulsar. The component 
widths and their separation are estimated from the lower boundary lines for the
distribution of widths with period. The emission is assumed to originate at a 
constant height above the surface for all the emission components. 
\label{fig_widper}}
\end{figure*}

\noindent
It has been shown observationally that the radio emission in normal pulsars 
originate at heights $R_E \sim$ 500 km \citep{bla91,von97,mit04,wel08,mit11}. 
It has also been shown in several cases that both the core and conal components
arise from similar heights within the pulsar magnetosphere \citep[see Table 2 
in][]{mit16a}. In addition we have now shown that both the core and conal 
components have similar LBL which shows a $P^{-0.5}$ dependence. Hence, the 
average emission beam in normal pulsars can be imagined to consist of a central
core component surrounded by conal components around similar heights as shown 
schematically in figure \ref{fig_widper}. It has been argued in earlier works 
\citep[e.g. R90,][]{mac12} that the $P^{-0.5}$ is an indication of the 
evolution of the components along the dipolar field lines since the open field 
line regions in dipolar magnetic fields are associated with the light cylinder 
radius which has a period dependence ($R_{LC} = Pc/2\pi$). 

The equation for the field lines in a dipolar field is given as :
\begin{equation}
r = R_C \sin^2\theta.
\label{eq4_1}
\end{equation}
Where $r$ and $\theta$ are the polar co-ordinates and $R_C$ is the constant of 
the field lines. 
The radio emission is along the tangential direction to the field lines given 
as :
\begin{equation}
\tan\phi = \frac{3\sin\theta\cos\theta}{3\cos^2\theta - 1},
\label{eq4_3}
\end{equation}
which for small angles can be expressed as $\phi = \frac{3}{2}\theta$. The 
components are bound by the dipolar field lines which follow the above equation
and the only period dependence is associated with the last open field lines 
where $R_C = R_{LC}$. So, if the emission height is constant for different 
periods ($r = R_E$) and the components occupy inner field lines, i.e. $R_C \neq
R_{LC}$, the width of the components should remain constant and not show any 
period dependence. 

The analysis of profile component widths reported in this work suggests a 
deviation from the conventional view that the period dependence of widths is 
a consequence of the dipolar geometry in pulsars. We have shown that the 
$P^{-0.5}$ dependence is indeed seen individually in the core components as 
well as the conal components. However, this dependence is not a result of 
dipolar geometry if we assume that the components occupy a small area in the 
open field line region of pulsars. The $P^{-0.5}$ dependence follows from the
light cylinder radius associated with the last open field line and hence any 
component with an edge not along the last open field line will not follow this 
dependence. The additional estimates of the underlying widths suggest that they
are identical in both the core and conal components. This indicates that the 
emission heights in normal pulsars are similar across the pulsar magnetosphere 
at any given frequency contrary to previous claims of the core and conal 
emission having different locations within the magnetosphere \citep{ran90,
ran93a}. 

\subsection{Period dependence of Profile widths}
\noindent
It is worthwhile to look at the implications of the component widths on the
overall profile width and their period dependence. Firstly, it is unlikely
that the measured full width at 50\% points stretches out to the last open 
field lines. As we discuss above the $P^{-0.5}$ scaling of overall widths is 
not likely due to the effect of dipolar geometry. However, to see the effect of
the LBL on full profile widths, specially in pulsars with more than one 
component, is difficult as the LBL is dominated by {\bf S$_t$} and {\bf S$_d$} 
classes with single components. As discussed in section~\ref{sec:intro} a 
$P^{-0.5}$ dependence is reported for the opening angle $\rho^{\nu}$ 
corresponding to measured widths for multiple component profiles ({\bf M} and 
{\bf T} classes). We argued that the dependence is an effect of the estimation 
process of $\rho^{\nu}$ and follows from the period dependence in the core 
widths. In order to preserve this relation we found that the quantity $F$ 
defined in eq.(\ref{eq5}) should be period independent. Based on our analysis 
we are now in a position to verify this claim. As shown earlier, the full width
at any frequency $\nu$ can be broken down as $W^{\nu} = W_{core}^{\nu} + 
2nW_{cone}^{\nu} + 2n\delta W^{\nu}$, with $n$=1,2 for the inner and outer cone
respectively. We have assumed identical widths for all the conal components 
(which is very likely given the presence of boundary line) as well as all the 
separation between components (less well constrained due to the absence of a 
clear boundary). We have now shown that both $W_{core}^{\nu}$ and 
$W_{cone}^{\nu}$ follow a $P^{-0.5}$ dependence individually and are more or 
less identical (which we call $W_{comp}^{\nu}$). The final part gives the 
distance between adjacent components and can be represented as 
$\delta W^{\nu} = W_{sep}^{\nu}-W_{comp}^{\nu}$, where $W_{sep}^{\nu}$ is the 
separation between the adjacent peaks of components. Our estimates of 
$W_{sep}^{\nu}$ in section~\ref{sec:res}, did not show an explicit period 
dependence as seen in the components. But boundary values $W_{sep}$ = 
2.66\degr~and $W_{comp}$ = 2.25\degr~implies $\delta W$ = 0.41\degr~; i.e 
$\delta W \approx 0.18W_{comp}$. Thus the distance between the components form 
a small fraction of the component widths. Hence, even if $\delta W$ do not show
a $P^{-0.5}$ dependence their contribution to the overall width will be small 
enough. This means that the factor $F$ in eq.(\ref{eq9}) can be approximated to
be largely period independent.

\section{Summary}
\noindent
In this work we have carried out detailed measurements to characterise the 
widths of the core and conal components separately for a large number of normal
pulsars. We have shown that the component widths when distributed with period
show the presence of identical lower boundary lines for both the core and 
conal components which follow a $P^{-0.5}$ dependence. These results firmly
establish the core and the conal components to be equivalent within the pulsar 
profile and eliminate any requirement for two different emission mechanisms 
for core and conal emission at different heights \citep{ran90,ran93a,mel17}. 
We have also highlighted that in normal pulsars where the radio emission is 
supposed to originate at heights of around 500 kilometers from the stellar 
surface throughout the pulsar profile, the $P^{-0.5}$ dependence of the lower 
boundary line is a result of the physical mechanism and not an outcome of the 
dipolar nature of the magnetic field in the emission region. In the absence of 
further constraints from observations more detailed modeling of the emission 
processes are required to explain the period dependence of components. \\\\

\noindent
{\bf Acknowledgments}: 
We thank the referee for the comments which helped to improve the paper. We 
thank Mihir Arjunwadkar for discussion on the LBL analysis techniques. We would
like to thank staff of Giant Meterwave Radio Telescope and National Center for 
Radio Astrophysics for providing valuable support in carrying out this project.
This work was supported by grants DEC-2012/05/B/ST9/03924, 
DEC-2013/09/B/ST9/02177 and UMO-2014/13/ B/ST9/00570 of the Polish National 
Science Centre.

\appendix




 
\footnotesize

\begin{longtable}{ccc | cccc | cccc}

\caption{THE TABLE SHOWS THE DETAILS OF THE MEASUREMENTS OF WIDTHS IN PULSARS. 
THE MEASUREMENTS INCLUDE WIDTHS OF COMPONENTS ($W_{50}$) AND SEPARATION BETWEEN
ADJACENT COMPONENTS ($W_{SEP}$). THE FIRST COLUMN PRESENTS THE NAME, THE SECOND
COLUMN THE PERIOD ($P$). COLUMNS 4-7 GIVE THE MEASUREMENTS AT 333~MHz, WHICH 
INCLUDE THE NUMBER OF COMPONENTS $N_{c}$, THE TECHNIQUE USED TO ESTIMATE THE 
COMPONENT (1 - INTEGRATED PROFILE, 2 - AVERAGING SUBPULSES, 3 - AVERAGING PEAKS
IN WINDOW) ALONG WITDH ESTIMATION METHOD (a - FULL WIDTH, b - HALF WIDTH, c - 
FITTING GAUSSIAN). COLUMNS 8-11 THE CORRESPONDING VALUES AT 618~MHz DATA. 
\label{tabela}}
\\

   &  &  & \multicolumn{4}{c}{333 MHz} & \multicolumn{4}{c}{618 MHz} \\

 No & PSR & $P$ & $N_{c}$ & Method  & $W_{50}$ & $W_{sep}$ & $N_{c}$ & Method & $W_{50}$ & $W_{sep}$ \\
 
  &  & [s]  &   &  & [deg] & [deg] &  &  & [deg] & [deg] \\ \hline

\endfirsthead
  &  &  & \multicolumn{4}{c}{333 MHz} & \multicolumn{4}{c}{618 MHz} \\

No & PSR & $P$ & $N_{c}$ & Method  & $W_{50}$ & $W_{sep}$ & $N_{c}$ & Method & $W_{50}$ & $W_{sep}$ \\ 
 
  &  & [s]  &   &  & [deg] & [deg] &  &  & [deg] & [deg] \\ \hline

\endhead
\endfoot
\hline
\endlastfoot

1 & B0031$-$07 & 0.94 & 2 & 3-a/b & \mysplit{11.29$\pm$0.13 \\ 12.27$\pm$0.13} & \mysplit{17.14$\pm$0.13} & 2 & 3-a/b & \mysplit{9.61$\pm$0.13 \\ 9.54$\pm$0.13} & \mysplit{16.21$\pm$0.13} \\
  &              &        &       &                 &                    &   &       &                  &                    &   \\         
2 & J0134$-$2937 & 0.14 & 1 & 1-a & \mysplit{16.20$\pm$0.60} & \mysplit{ - } & 1 & 1-a & \mysplit{18.80$\pm$0.90} & \mysplit{ - } \\ 
  &              &        &       &                 &                    &   &       &                  &                    &   \\         
3 & B0148$-$06 & 1.46 & 2 & 1-a & \mysplit{6.89$\pm$0.09 \\ 4.88$\pm$0.09} & \mysplit{30.51$\pm$0.09} & 2 & 1-a & \mysplit{6.49$\pm$0.09 \\ 6.33$\pm$0.09} & \mysplit{27.01$\pm$0.09} \\
  &              &        &       &                 &                    &   &       &                  &                    &   \\         
4 & B0149$-$16 & 0.83 & 2 & 1-a,b & \mysplit{3.38$\pm$0.15 \\ 5.01$\pm$0.15} & \mysplit{5.72$\pm$0.15} & 2 & 1-a,b & \mysplit{2.75$\pm$0.15 \\ 4.81$\pm$0.15} & \mysplit{5.36$\pm$0.15}  \\ 
  &              &        &       &                 &                    &   &       &                  &                    &   \\         
5 & B0203$-$40 & 0.63 & 2 & 1-a,c & \mysplit{3.83$\pm$0.20 \\ 3.33$\pm$0.20} & \mysplit{6.46$\pm$0.20} & 1 & 1-a & \mysplit{4.13$\pm$1.00} & \mysplit{ - } \\ 
  &              &        &       &                 &                    &   &       &                  &                    &   \\         
6 & B0301$-$19 & 1.39 & 2 & 1-a & \mysplit{3.20$\pm$0.09 \\ 4.41$\pm$0.09} & \mysplit{12.69$\pm$0.09} & 2 & 1-a & \mysplit{4.75$\pm$0.09 \\ 4.71$\pm$0.09} & \mysplit{10.27$\pm$0.09} \\
  &              &        &       &                 &                    &   &       &                  &                    &   \\         
7 & B0450$-$18 & 0.55 & 4 & 1/3-a,b,b & \mysplit{4.49$\pm$0.23 \\ 6.49$\pm$0.23 \\ - \\ 5.03$\pm$0.23} & \mysplit{7.90$\pm$0.23} & 4 & 1-b,c,b & \mysplit{4.03$\pm$0.23 \\ 8.46$\pm$0.23 \\ - \\ 6.10$\pm$0.23} & \mysplit{7.58$\pm$0.23} \\ 
  &              &        &       &                 &                    &   &       &                  &                    &   \\         
8 & B0523+11 & 0.35 & 2 & 1-a,b & \mysplit{4.00$\pm$0.35 \\ 4.79$\pm$0.35} & \mysplit{12.24$\pm$0.35} & 2 & 1-a,a & \mysplit{7.83$\pm$0.35 \\ 6.51$\pm$0.35} & \mysplit{11.24$\pm$0.35} \\ 
  &              &        &       &                 &                    &   &       &                  &                    &   \\ 
9 & B0525+21 & 3.75 & 2 & 1-a,a & \mysplit{2.78$\pm$0.03 \\ 2.55$\pm$0.03} & \mysplit{14.69$\pm$0.03} & - & - & \mysplit{ - } & \mysplit{ - } \\ 
  &              &        &       &                 &                    &   &       &                  &                    &   \\         
10 & B0540+23 & 0.25 &  2 & 1/3-a & \mysplit{8.34$\pm$0.51 \\ - } & \mysplit{18.72$\pm$0.51} & 2 & 1/3-a & \mysplit{7.40$\pm$0.51 \\ - } & \mysplit{10.44$\pm$0.51} \\ 
  &              &        &       &                 &                    &   &       &                  &                    &   \\         
11 & B0611+22 & 0.33 &  1 & 1-a & \mysplit{6.95$\pm$0.37} & \mysplit{ - } & 1 & 1-a & \mysplit{5.91$\pm$0.38} & \mysplit{ - } \\ 
  &              &        &       &                 &                    &   &       &                  &                    &   \\         
12 & B0626+24 & 0.48 & 2 & 1-a,b & \mysplit{5.86$\pm$0.26 \\ 6.68$\pm$0.26} & \mysplit{8.17$\pm$0.26} & 2 & 1-a & \mysplit{6.36$\pm$0.26 \\ - } & \mysplit{6.13$\pm$0.26} \\ 
  &              &        &       &                 &                    &   &       &                  &                    &   \\         
13 & B0628$-$28 & 1.24 & 1 & 1-a & \mysplit{19.00$\pm$0.10} & \mysplit{ - } & 1 & 1-a & \mysplit{18.33$\pm$0.10} & \mysplit{ - } \\ 
  &              &        &       &                 &                    &   &       &                  &                    &   \\         
14 & B0656+14 & 0.38 &  - & 1-a & \mysplit{15.60$\pm$0.30} & \mysplit{ - } & - & 1-a & \mysplit{14.30$\pm$0.60} & \mysplit{ - } \\ 
  &              &        &       &                 &                    &   &       &                  &                    &   \\         
15 & B0727$-$18 & 0.51 &  3 & 1-a,b & \mysplit{3.17$\pm$0.25 \\ - \\ 7.23$\pm$0.25} & \mysplit{ - } & 3 & 1-a,b & \mysplit{3.38$\pm$0.25 \\ - \\ 6.33$\pm$0.25} & \mysplit{ - } \\ 
  &              &        &       &                 &                    &   &       &                  &                    &   \\         
16 & B0736$-$40 & 0.37 & -  & - & \mysplit{ - } & \mysplit{ - } & 3 & 1-b,b & \mysplit{ - \\ 7.99$\pm$0.33 \\ 21.5$\pm$0.33} & \mysplit{ - \\ 17.72$\pm$0.33} \\ 
  &              &        &       &                 &                    &   &       &                  &                    &   \\         
17 & B0740$-$28 & 0.17 &  3 & 1 -b,b & \mysplit{6.40$\pm$0.75 \\ - \\ 7.35$\pm$0.75} & \mysplit{5.74$\pm$0.75} & 4 & 1-b,b & \mysplit{5.39$\pm$0.75 \\ - \\ 4.28$\pm$0.75} & \mysplit{2.66$\pm$0.75 \\ 3.75$\pm$0.75 \\ - } \\ 
  &              &        &       &                 &                    &   &       &                  &                    &   \\         
18 & B0756$-$15 & 0.68 & 2 & 1-b,b & \mysplit{3.95$\pm$0.18 \\ 2.16$\pm$0.19} & \mysplit{2.08$\pm$0.18} & 2 & 1-b,b & \mysplit{2.68$\pm$0.18 \\ 2.15$\pm$0.18} & \mysplit{2.01$\pm$0.18} \\ 
  &              &        &       &                 &                    &   &       &                  &                    &   \\         
19 & B0818$-$13 & 1.24 & 2 & 3-a/b & \mysplit{3.61$\pm$0.10 \\ 3.44$\pm$0.10} & \mysplit{2.64$\pm$0.10} & 2 & 3-a/b & \mysplit{3.60$\pm$0.1 \\ 2.45$\pm$0.10} & \mysplit{3.07$\pm$0.10} \\ 
  &              &        &       &                 &                    &   &       &                  &                    &   \\         
20 & B0818$-$41 & 0.55 & - & - & \mysplit{ - } & \mysplit{-} & 2 & 1-b,b & \mysplit{49.94$\pm$ 2.33 \\ 50.76$\pm$2.33 } & \mysplit{72.00$\pm$2.32} \\ 
  &              &        &       &                 &                    &   &       &                  &                    &   \\         
21 & B0834+06 & 1.27 & 2 & 1-a,b & \mysplit{2.87$\pm$0.10 \\ 3.08$\pm$0.10} & \mysplit{4.59$\pm$0.10} & 2 & 1-a,b & \mysplit{2.86$\pm$0.1 \\ 2.83$\pm$0.10} & \mysplit{4.93$\pm$0.10} \\ 
  &              &        &       &                 &                    &   &       &                  &                    &   \\         
22 & B0844$-$35 & 1.12 & 4/5 & 1-b,a,b & \mysplit{6.83$\pm$0.56 \\ - \\ - \\ 5.18$\pm$0.56 \\ 4.26$\pm$0.56} & \mysplit{ - \\ - \\ 6.75$\pm$0.56 \\ 5.16$\pm$0.56} & 4/5 & 1-b,a,b & \mysplit{5.64$\pm$0.57 \\ - \\ - \\ 4.29$\pm$0.56 \\ 2.83$\pm$0.56} & \mysplit{ - \\ - \\ 5.56$\pm$0.56 \\ 4.76$\pm$0.56} \\ 
  &              &        &       &                 &                    &   &       &                  &                    &   \\         
23 & J0905$-$4536 & 0.99 & - & - & \mysplit{ - } & \mysplit{ - } & 2 & 1-a,a & \mysplit{12.87$\pm$1.92	\\ 19.25$\pm$1.93} & \mysplit{ 48.54$\pm$1.91} \\ 
  &              &        &       &                 &                    &   &       &                  &                    &   \\         
24 & J0905$-$5127 & 0.35 &  1 & 1/2-b,b & \mysplit{9.73$\pm$0.36 } & \mysplit{ - } & 2 & 1/2-b,b & \mysplit{4.37$\pm$0.36 \\ 2.77$\pm$0.36 } & \mysplit{4.35$\pm$0.36 } \\ 
  &              &        &       &                 &                    &   &       &                  &                    &   \\         
25 & B0906$-$17 & 0.40 & 2 & 1-a & \mysplit{8.05$\pm$0.31 \\ - } & \mysplit{9.59$\pm$0.31 } & 2 & 1-a & \mysplit{8.14$\pm$0.31 \\ - } & \mysplit{ - } \\ 
  &              &        &       &                 &                    &   &       &                  &                    &   \\         
26 & B0919+06 & 0.43 &  3 & 1-b,b & \mysplit{ - \\ 5.86$\pm$0.29 \\ 5.36$\pm$0.29} & \mysplit{ - \\	3.29$\pm$0.29 } & 3 & 1-b,b & \mysplit{ - \\ 4.25$\pm$0.29 \\ 5.33$\pm$0.29 } & \mysplit{ - \\ 2.21$\pm$0.29 } \\ 
  &              &        &       &                 &                    &   &       &                  &                    &   \\         
27 & B0942$-$13 & 0.57 &  2 & 1/3-b & \mysplit{2.95$\pm$0.22 \\ - } & \mysplit{ - } & 2 & 1/3-b,b & \mysplit{ 2.41$\pm$0.22 \\ 4.11$\pm$0.22 } & \mysplit{1.24$\pm$0.22 } \\ 
  &              &        &       &                 &                    &   &       &                  &                    &   \\         
28 & B0950+08 & 0.25 &  2 & 1/3-a,a & \mysplit{16.06$\pm$0.50 \\ 14.57$\pm$0.50 } & \mysplit{34.36$\pm$0.50 } & 2 & 1/3-a,a & \mysplit{13.50$\pm$0.50 \\ 11.97$\pm$0.50} & \mysplit{33.63$\pm$0.50 } \\ 
  &              &        &       &                 &                    &   &       &                  &                    &   \\         
29 & B0957$-$47 & 0.67 &  - & - & \mysplit{ - } & \mysplit{ - } & 2 & 1-a,a & \mysplit{18.93$\pm$1.88 \\ 10.48$\pm$1.88 } & \mysplit{45.17$\pm$1.88} \\ 
  &              &        &       &                 &                    &   &       &                  &                    &   \\         
30 & J1034$-$3224 & 1.15 &  5 & 1-a,b,b,b & \mysplit{ - \\ 6.52$\pm$0.11 \\ 13.24$\pm$0.11 \\ - \\ 9.16$\pm$0.11 \\ 7.49$\pm$0.11} & \mysplit{ - \\ 13.77$\pm$0.11 \\ - \\	- \\ 7.31$\pm$0.11} & 6 & 1-a,a,c,c,b & \mysplit{18.82$\pm$1.09 \\ 7.50$\pm$1.09 \\ 12.65$\pm$1.09 \\ - \\ 11.81$\pm$1.09 \\ 7.1$\pm$1.09 } & \mysplit{30.84$\pm$1.09 \\ 13.88$\pm$1.09 \\ - \\ - \\ 7.71$\pm$1.09} \\ 
  &              &        &       &                 &                    &   &       &                  &                    &   \\         
31 & B1039$-$19 & 1.39 &  - & - & \mysplit{ - } & \mysplit{ - } & 3 & 1-a,b,a & \mysplit{2.47$\pm$0.09 \\ 6.58$\pm$0.09 \\ 5.74$\pm$0.09 } & \mysplit{2.87$\pm$0.09 \\ 8.87$\pm$0.09} \\ 
  &              &        &       &                 &                    &   &       &                  &                    &   \\         
32 & B1114$-$41 & 0.94 &  1 & 1-a & \mysplit{5.70$\pm$0.20 } & \mysplit{ - } & 1 & 1-a & \mysplit{5.30$\pm$0.20 } & \mysplit{ - } \\ 
  &              &        &       &                 &                    &   &       &                  &                    &   \\         
33 & B1133+16 & 1.19 &  2 & 1-a,a & \mysplit{1.95$\pm$0.11 \\ 3.99$\pm$0.11 } & \mysplit{7.60$\pm$0.11 } & 2 & 1-a,a & \mysplit{1.62$\pm$0.11 \\ 4.22$\pm$0.11 } & \mysplit{6.11$\pm$0.11  } \\ 
  &              &        &       &                 &                    &   &       &                  &                    &   \\         
34 & B1237+25 & 1.38 &  5 & 1-a,b,a,c,b & \mysplit{1.51$\pm$0.09 \\ 1.60$\pm$0.09 \\ 2.01$\pm$0.09 \\ 2.24$\pm$0.09 \\ 1.76$\pm$0.09 } & \mysplit{2.56$\pm$0.09 \\ 4.48$\pm$0.09 \\ 2.82$\pm$0.09 \\ 2.24$\pm$0.09 } & 5 & 1-a,b,b,b & \mysplit{1.40$\pm$0.09 \\ 1.48$\pm$0.09 \\ 3.06$\pm$0.09 \\ - \\ 2.2$\pm$0.09} & \mysplit{2.43$\pm$0.09 \\ 4.16$\pm$0.09 \\ 2.50$\pm$0.09 \\ 1.66$\pm$0.09 } \\ 
  &              &        &       &                 &                    &   &       &                  &                    &   \\         
35 & B1254$-$10 & 0.62 &  2 & 1-a,a & \mysplit{2.95$\pm$0.20 \\ 3.08$\pm$0.20 } & \mysplit{10.61$\pm$0.20 } & 2 & 1-a,a & \mysplit{4.10$\pm$0.21 \\ 2.77$\pm$0.20} & \mysplit{10.61$\pm$0.20 } \\ 
  &              &        &       &                 &                    &   &       &                  &                    &   \\         
36 & B1322+83 & 0.67 &  - & - & \mysplit{ - } & \mysplit{ - } & 1 & 1-a & \mysplit{2.23$\pm$0.19 } & \mysplit{ - } \\ 
  &              &        &       &                 &                    &   &       &                  &                    &   \\         
37 & B1325$-$43 & 0.53 &  2 & 1-b,b & \mysplit{4.59$\pm$0.24 \\ 6.31$\pm$0.24 } & \mysplit{6.81$\pm$0.24 } & 1 & 1-a & \mysplit{8.83$\pm$0.24 } & \mysplit{ - } \\ 
  &              &        &       &                 &                    &   &       &                  &                    &   \\         
38 & B1325$-$49 & 1.48 &  5 & 1-b,b,b,a & \mysplit{2.51$\pm$0.09 \\ - \\ 1.90$\pm$0.09 \\ 2.44$\pm$0.08 \\ 1.37$\pm$0.08} & \mysplit{2.45$\pm$0.09 \\ 4.54$\pm$0.09 \\ 3.71$\pm$0.09 \\ 2.15$\pm$0.09 } & 5 & 1-b,b,b,a & \mysplit{2.33$\pm$0.09 \\ - \\ 1.81$\pm$0.09 \\ 2.02$\pm$0.09 \\ 1.53$\pm$0.08 } & \mysplit{2.33$\pm$0.09 \\ 3.83$\pm$0.09 \\ 3.53$\pm$0.09 \\ 2.39$\pm$0.09 } \\ 
  &              &        &       &                 &                    &   &       &                  &                    &   \\         
39 & J1418$-$3921 & 1.10 &  -  & - & \mysplit{ - } & \mysplit{ - } & 2 & 1-b & \mysplit{4.52$\pm$0.57 \\ - } & \mysplit{ - } \\ 
  &              &        &       &                 &                    &   &       &                  &                    &   \\         
40 & B1504$-$43 & 0.29 &  1 & 1-a & \mysplit{7.40$\pm$0.80 } & \mysplit{ - } & 1 & 1-a & \mysplit{6.20$\pm$0.80 } & \mysplit{ - } \\ 
  &              &        &       &                 &                    &   &       &                  &                    &   \\         
41 & B1524$-$39 & 2.42 &  2 & 1-a,a & \mysplit{1.89$\pm$0.05 \\ 2.96$\pm$0.05} & \mysplit{3.00$\pm$0.05 } & 2 & 1-a,a & \mysplit{2.20$\pm$0.26 \\ 3.09$\pm$0.26 } & \mysplit{2.75$\pm$0.26 } \\
  &              &        &       &                 &                    &   &       &                  &                    &   \\         
42 & J1549$-$4848 & 0.29 & 1 & 1-a & \mysplit{6.40$\pm$0.80 } & \mysplit{ - } & 1 & 1-a & \mysplit{7.10$\pm$0.80 } & \mysplit{ - } \\ 
  &              &        &       &                 &                    &   &       &                  &                    &   \\         
43 & B1552$-$31 & 0.52 &  2 & 1-a,a & \mysplit{4.84$\pm$0.24 \\ 5.05$\pm$0.24 } & \mysplit{15.73$\pm$0.24 } & 2 & 1-a,a & \mysplit{4.63$\pm$0.24 \\ 5.46$\pm$0.24 } & \mysplit{14.7$\pm$0.24 } \\ 
  &              &        &       &                 &                    &   &       &                  &                    &   \\         
44 & J1557$-$4258 & 0.33 &  - & - & \mysplit{ - } & \mysplit{ - } & 3 & 1-c,a,c & \mysplit{4.17$\pm$0.41 \\ 4.33$\pm$0.38 \\ 7.12$\pm$0.40 } & \mysplit{6.99$\pm$0.38 \\ 9.69$\pm$0.38 } \\ 
  &              &        &       &                 &                    &   &       &                  &                    &   \\         
45 & B1556$-$44 & 0.26 &  3 & 1/3-b,a & \mysplit{10.76$\pm$0.49 \\ 8.04$\pm$0.49 \\ - } & \mysplit{13.79$\pm$0.49 \\ 6.9$\pm$0.49} & 2/ 3 & 1-b,a & \mysplit{6.49$\pm$0.49 \\ 8.12$\pm$0.49 \\ - } & \mysplit{13.10$\pm$0.49 } \\ 
  &              &        &       &                 &                    &   &       &                  &                    &   \\         
46 & B1558$-$50 & 0.86 &  - & - & \mysplit{ - } & \mysplit{ - } & 2 & 1-a,b & \mysplit{2.73$\pm$0.14 \\ 3.77$\pm$0.14} & \mysplit{4.51$\pm$0.14 } \\ 
  &              &        &       &                 &                    &   &       &                  &                    &   \\         
47 & J1603$-$2531 & 0.28 &  1 & 1-a & \mysplit{10.60$\pm$0.80 } & \mysplit{ - } & 1 & 1-a & \mysplit{8.80$\pm$0.80 } & \mysplit{ - } \\ 
  &              &        &       &                 &                    &   &       &                  &                    &   \\         
48 & B1600$-$49 & 0.33 &  - & - & \mysplit{ - } & \mysplit{ - } & 3 & 1-c,a,c & \mysplit{3.52$\pm$0.39 \\ 4.88$\pm$0.38 \\ 5.25$\pm$0.42} & \mysplit{7.30$\pm$0.39 \\ 8.20$\pm$0.38 } \\ 
  &              &        &       &                 &                    &   &       &                  &                    &   \\         
49 & J1625$-$4048 & 2.36 &  - & - & \mysplit{ - } & \mysplit{ - } & 3 & 1-a,a,a & \mysplit{1.73$\pm$0.05 \\ 3.02$\pm$0.05 \\ 3.96$\pm$0.06 } & \mysplit{10.15$\pm$0.05 \\ 6.76$\pm$0.05 } \\ 
  &              &        &       &                 &                    &   &       &                  &                    &   \\         
50 & B1634$-$45 & 0.12 &  1 & - & \mysplit{ - } & \mysplit{ - } & 1 & 1-a & \mysplit{16.40$\pm$1.90 } & \mysplit{ - } \\ 
  &              &        &       &                 &                    &   &       &                  &                    &   \\         
51 & B1642$-$03 & 0.39 &  - & - & \mysplit{ - } & \mysplit{ - } & 3 & 1/3-a,b & \mysplit{ - \\ 3.63$\pm$0.32 \\ 5.60$\pm$0.32 } & \mysplit{5.48$\pm$0.32 \\ 8.68$\pm$0.32 } \\ 
  &              &        &       &                 &                    &   &       &                  &                    &   \\         
52 & J1648$-$3256 & 0.72 &  1 & 1-a & \mysplit{5.98$\pm$0.52 } & \mysplit{ - } & 1 & 1-a & \mysplit{5.71$\pm$0.52} & \mysplit{ - } \\ 
  &              &        &       &                 &                    &   &       &                  &                    &   \\         
53 & J1700$-$3312 & 1.36 &  3 & 1-b,b & \mysplit{4.32$\pm$0.09 \\ - \\ 4.58$\pm$0.10 } & \mysplit{-} & 3 & 1-b,b & \mysplit{2.01$\pm$0.09 \\ - \\ 3.28$\pm$0.09 } & \mysplit{3.58$\pm$0.09 \\ 5.8$\pm$0.09} \\ 
  &              &        &       &                 &                    &   &       &                  &                    &   \\         
54 & B1700$-$32 & 1.21 &  3 & 1-b,c,b & \mysplit{5.05$\pm$0.10 \\ 4.80$\pm$0.10 \\ 3.52$\pm$0.10 } & \mysplit{5.40$\pm$0.10 \\ 5.03$\pm$0.10 } & 3 & 1-b,c,b & \mysplit{4.61$\pm$0.10 \\ 5.76$\pm$0.10 \\ 2.74$\pm$0.10 } & \mysplit{4.78$\pm$0.10 \\ 4.57$\pm$0.10} \\ 
  &              &        &       &                 &                    &   &       &                  &                    &   \\         
55 & J1705$-$3423 & 0.26 &  - & - & \mysplit{ - } & \mysplit{ - } & 1 & 1-a & \mysplit{20.09$\pm$0.49 } & \mysplit{ - } \\ 
  &              &        &       &                 &                    &   &       &                  &                    &   \\         
56 & B1706$-$16 & 0.65 &  1 & 1-a & \mysplit{5.80$\pm$0.19 } & \mysplit{ - } & 1 & 1-a & \mysplit{5.57$\pm$0.19 } & \mysplit{ - } \\ 
  &              &        &       &                 &                    &   &       &                  &                    &   \\         
57 & B1706$-$44 & 0.10 &  1 & 1-a & \mysplit{42.30$\pm$ 2.20 } & \mysplit{ - } & 1 & 1-a & \mysplit{31.90$\pm$ 2.20 } & \mysplit{ - } \\ 
  &              &        &       &                 &                    &   &       &                  &                    &   \\         
58 & B1717$-$29 & 0.62 &  3 & 3	a/b & \mysplit{4.99$\pm$0.21 \\ 4.99$\pm$0.20 \\ 4.99$\pm$0.20 } & \mysplit{6.53$\pm$0.20 \\ 10.20$\pm$0.20 } & 3 & 3-a/b & \mysplit{4.24$\pm$0.20 \\ 4.24$\pm$0.20 \\ 4.24$\pm$0.20 } & \mysplit{7.56$\pm$0.20 \\ 8.70$\pm$0.20 } \\ 
  &              &        &       &                 &                    &   &       &                  &                    &   \\         
59 & B1718$-$32 & 0.48 &  - & - & \mysplit{ - } & \mysplit{ - } & 2 & 1-b,b & \mysplit{3.81$\pm$0.26 \\ 7.81$\pm$0.26 } & \mysplit{5.57$\pm$0.26 }\\ 
  &              &        &       &                 &                    &   &       &                  &                    &   \\         
60 & B1719$-$37 & 0.24 &   1 & 1-a & \mysplit{11.85$\pm$1.60 } & \mysplit{ - } & 1 & 1-a & \mysplit{7.52$\pm$1.60 } & \mysplit{ - } \\ 
  &              &        &       &                 &                    &   &       &                  &                    &   \\         
61 & J1727$-$2739 & 1.29 &  2 & 1	a & \mysplit{13.51$\pm$1.50 \\ 10.04$\pm$1.47 } & \mysplit{25.79$\pm$1.46} & 2 & 1-a & \mysplit{6.51$\pm$1.47 \\ 6.30$\pm$1.77 } & \mysplit{22.69$\pm$1.46 } \\ 
  &              &        &       &                 &                    &   &       &                  &                    &   \\         
62 & B1727$-$47 & 0.83 &   3 & 1-b,b & \mysplit{2.71$\pm$0.15 \\ - \\ 2.87$\pm$0.15 } & \mysplit{ - } & 3 & 1-a,b & \mysplit{2.50$\pm$0.15 \\ - \\ 2.37$\pm$0.15 } & \mysplit{ - } \\ 
  &              &        &       &                 &                    &   &       &                  &                    &   \\         
63 & B1730$-$22 & 0.87 &  4 & 1-a,c,a & \mysplit{ - \\ 5.13$\pm$0.14 \\ 6.02$\pm$0.14 \\ 11.65$\pm$0.14 } & \mysplit{ - \\ 10.46$\pm$0.14 \\ 11.07$\pm$0.14 } & 4 & 1-a,c,a & \mysplit{ - \\ 6.65$\pm$0.29 \\ 10.07$\pm$0.29 \\ 4.52$\pm$0.29 } & \mysplit{ - \\ 10.05$\pm$0.29 \\ 10.67$\pm$0.29 } \\ 
  &              &        &       &                 &                    &   &       &                  &                    &   \\         
64 & B1730$-$37 & 0.34 &  - & - & \mysplit{ - } & \mysplit{ - } & 2 & 1-a,a & \mysplit{9.16$\pm$1.88 \\ 14.95$\pm$1.91 } & \mysplit{40.88$\pm$1.86 } \\ 
  &              &        &       &                 &                    &   &       &                  &                    &   \\         
65 & B1732$-$07 & 0.42 &  3 & 1-b,a,b & \mysplit{4.31$\pm$0.30 \\ 3.83$\pm$0.30 \\ 3.63$\pm$0.30 } & \mysplit{9.71$\pm$0.30 \\ 5.06$\pm$0.30} & 3 & 1-b,a,b & \mysplit{ - \\ 3.69$\pm$0.30 \\ 4.37$\pm$0.30 } & \mysplit{9.28$\pm$0.30 \\ 4.86$\pm$0.30 } \\ 
  &              &        &       &                 &                    &   &       &                  &                    &   \\         
66 & B1736$-$29 & 0.32 &   - & - & \mysplit{ - } & \mysplit{ - } & 1 & 1-a & \mysplit{8.16$\pm$0.40 } & \mysplit{ - } \\ 
  &              &        &       &                 &                    &   &       &                  &                    &   \\         
67 & B1737+13 & 0.80 &  5 & 1/3-a,c,b,b & \mysplit{2.79$\pm$0.16 \\ 4.15$\pm$0.16 \\ 3.81$\pm$0.16 \\ - \\ 3.18$\pm$0.16 } & \mysplit{3.74$\pm$0.16 \\ 6.94$\pm$0.16 \\ 3.49$\pm$0.16 \\ 4.96$\pm$0.16 } & 5 & 1/3-a,c,b,b & \mysplit{2.89$\pm$0.16	\\ 3.94$\pm$0.16 \\ 4.84$\pm$0.16 \\ - \\ 3.01$\pm$0.16} & \mysplit{3.63$\pm$0.16 \\ 5.73$\pm$0.16 \\ - \\ - } \\ 
  &              &        &       &                 &                    &   &       &                  &                    &   \\         
68 & B1738$-$08 & 2.04 &  4 & 1-b,b & \mysplit{2.72$\pm$0.06 \\ - \\ - \\ 3.48$\pm$0.06} & \mysplit{3.55$\pm$0.06 \\ 8.24$\pm$0.06 \\ 2.60$\pm$0.06 } & 4 & 1-b,b & \mysplit{3.42$\pm$0.06 \\ - \\ - \\ 2.70$\pm$0.06 } & \mysplit{3.16$\pm$0.06 \\ 6.11$\pm$0.06 \\ 3.08$\pm$0.06 } \\ 
  &              &        &       &                 &                    &   &       &                  &                    &   \\         
69 & B1737$-$39 & 0.51 &  2 & 1-b & \mysplit{10.88$\pm$0.24 \\ - } & \mysplit{ - } & 2 & 1-a & \mysplit{6.51$\pm$0.24 \\ - } & \mysplit{ - } \\ 
  &              &        &       &                 &                    &   &       &                  &                    &   \\         
70 & B1742$-$30 & 0.37 &   4 & 1-a,a & \mysplit{5.86$\pm$0.35	\\ - \\ 8.29$\pm$0.34 \\ -  } & \mysplit{ - } & 4 & 1/2/3-a,b,a & \mysplit{1.71$\pm$0.34 \\ 4.28$\pm$0.34 \\ 3.46$\pm$0.34 \\ - } & \mysplit{10.83$\pm$0.34 \\ 3.88$\pm$0.34 \\ 11.33$\pm$0.34 } \\ 
  &              &        &       &                 &                    &   &       &                  &                    &   \\         
71 & B1745$-$12 & 0.39 &  3 & 1-b,b,c & \mysplit{3.62$\pm$0.32 \\ 6.27$\pm$0.32 \\ 9.51$\pm$0.33 } & \mysplit{4.49$\pm$0.32 \\ 10.66$\pm$0.32 } & 3 & 1-b,b,c & \mysplit{3.28$\pm$0.32 \\ 5.77$\pm$0.32 \\ 7.14$\pm$0.32 } & \mysplit{4.27$\pm$0.32 \\ 5.85$\pm$0.32 } \\ 
  &              &        &       &                 &                    &   &       &                  &                    &   \\         
72 & J1750$-$3503 & 0.68 &  1 & 1-a & \mysplit{38.30$\pm$0.30 } & \mysplit{ - } & 1 & 1-a & \mysplit{34.00$\pm$0.30 } & \mysplit{ - } \\ 
  &              &        &       &                 &                    &   &       &                  &                    &   \\         
73 & B1747$-$46 & 0.74 &  2 & 1-b,b & \mysplit{3.53$\pm$0.17 \\ 4.44$\pm$0.17 } & \mysplit{4.17$\pm$0.17 } & 2 & 1-b,b & \mysplit{3.08$\pm$0.17 \\ 4.98$\pm$0.17 } & \mysplit{3.70$\pm$0.17 } \\ 
  &              &        &       &                 &                    &   &       &                  &                    &   \\         
74 & B1749$-$28 & 0.56 &  3 & 1/3-b,b & \mysplit{3.23$\pm$0.22 \\ 3.40$\pm$0.22 \\ - } & \mysplit{8.03$\pm$0.22 } & 3 & 1/3-b,b,b & \mysplit{2.71$\pm$0.22 \\ 3.10$\pm$0.22 \\ 3.17$\pm$0.22 } & \mysplit{8.74$\pm$0.22 \\ 1.10$\pm$0.22 } \\ 
  &              &        &       &                 &                    &   &       &                  &                    &   \\         
75 & B1754$-$24 & 0.23 &   - & - & \mysplit{ - } & \mysplit{ - } & 1 & 1-a & \mysplit{22.30$\pm$1.00 } & \mysplit{ - } \\ 
  &              &        &       &                 &                    &   &       &                  &                    &   \\         
76 & B1758$-$03 & 0.92 &  1 /3 & 1-a & \mysplit{ - \\ 4.73$\pm$0.14 \\ - } & \mysplit{ - } & 3 & 1-a & \mysplit{ - \\ 2.83$\pm$0.14 \\ - } & \mysplit{ - } \\ 
  &              &        &       &                 &                    &   &       &                  &                    &   \\         
77 & B1758$-$29 & 1.08 &  3 & 1-a,a,a & \mysplit{3.23$\pm$0.35 \\ 5.46$\pm$0.35 \\ 3.73$\pm$0.35 } & \mysplit{9.33$\pm$0.35 \\ 8.84$\pm$0.35 } & 3 & 1-a,a,a & \mysplit{2.71$\pm$0.35 \\ 6.70$\pm$0.35 \\ 5.78$\pm$0.35 } & \mysplit{7.86$\pm$0.35 \\ 8.59$\pm$0.35} \\ 
  &              &        &       &                 &                    &   &       &                  &                    &   \\         
78 & B1804$-$08 & 0.16 &  3 & 1-a & \mysplit{ - \\ 10.87$\pm$0.77 \\ - } & \mysplit{ - } & 3 & 1-a & \mysplit{ - \\ 5.77$\pm$0.77 \\ - } & \mysplit{-} \\ 
  &              &        &       &                 &                    &   &       &                  &                    &   \\         
79 & J1808$-$0813 & 0.88 &  1 & 1-a & \mysplit{14.90$\pm$0.30 } & \mysplit{ - } & 1 & 1-a & \mysplit{11.20$\pm$0.30 } & \mysplit{ - } \\ 
  &              &        &       &                 &                    &   &       &                  &                    &   \\         
80 & B1813$-$26 & 0.59 &  2 & 1-a,a & \mysplit{11.12$\pm$1.06 \\ 11.47$\pm$1.06 } & \mysplit{26.20$\pm$1.06 } & 2 & 1-a,b & \mysplit{10.99$\pm$1.06 \\ 8.23$\pm$1.07 } & \mysplit{21.70$\pm$1.06 } \\ 
  &              &        &       &                 &                    &   &       &                  &                    &   \\         
81 & B1813$-$36 & 0.39 &  3 & 1/3-a & \mysplit{ - \\ 5.82$\pm$0.32 \\ - } & \mysplit{5.70$\pm$0.32 } & - & - & \mysplit{ - } & \mysplit{ - } \\ 
  &              &        &       &                 &                    &   &       &                  &                    &   \\         
82 & J1817$-$3837 & 0.38 &  - & - & \mysplit{ - } & \mysplit{ - } & 3 & 1-b & \mysplit{2.57$\pm$0.60 \\ - \\ - } & \mysplit{ - } \\ 
  &              &        &       &                 &                    &   &       &                  &                    &   \\         
83 & B1818$-$04 & 0.60 &  2 & 1-b & \mysplit{4.75$\pm$0.21 \\	- } & \mysplit{ - } & 1/2 & 1-a & \mysplit{4.74$\pm$0.21 \\ - } & \mysplit{ - } \\ 
  &              &        &       &                 &                    &   &       &                  &                    &   \\         
84 & B1819$-$22 & 1.87 &  2 & 3-b/a & \mysplit{6.13$\pm$0.20\\ 6.66$\pm$0.20 } & \mysplit{8.60$\pm$0.20} & 2 & 3-b /a & \mysplit{4.59$\pm$0.20 \\ 4.78$\pm$0.20 } & \mysplit{7.82$\pm$0.20 \\ } \\ 
  &              &        &       &                 &                    &   &       &                  &                    &   \\         
85 & J1823$-$0154 & 0.76 &  2 & 1-a & \mysplit{ - \\ 60$\pm$0.30 } & \mysplit{ - } & 2 & 1-b,a & \mysplit{2.80$\pm$0.16 \\ 2.31$\pm$0.16 } & \mysplit{2.56$\pm$0.16 } \\ 
  &              &        &       &                 &                    &   &       &                  &                    &   \\         
86 & B1821+05 & 0.75 &  3 & 1-a,a,b & \mysplit{3.10$\pm$0.17 \\ 4.11$\pm$0.17 \\ 6.19$\pm$0.17 } & \mysplit{13.99$\pm$0.17 \\ 10.46$\pm$0.17 } & 3 & 1-a,a & \mysplit{2.88$\pm$0.17 \\ 5.48$\pm$0.17 \\ - } & \mysplit{13.40$\pm$0.17 \\ 7.90$\pm$0.17 } \\  
  &              &        &       &                 &                    &   &       &                  &                    &   \\         
87 & B1820$-$31 & 0.28 &   2 & 1/3-a & \mysplit{ - \\ 6.40$\pm$0.44} & \mysplit{4.06$\pm$0.44 } & 2 & 1/3-a & \mysplit{ - \\ 6.50$\pm$0.44 } & \mysplit{3.74$\pm$0.44 } \\ 
  &              &        &       &                 &                    &   &       &                  &                    &   \\         
88 & B1831$-$04 & 0.29 &  5 & 1-a,a,a,a & \mysplit{16.07$\pm$1.30 \\ - \\ 21.23$\pm$1.30 \\ 14.33$\pm$1.30 \\ 14.00$\pm$1.30} & \mysplit{20.20$\pm$1.30 \\ 21.12$\pm$1.30 \\ 35.81$\pm$1.30 \\ 22.04$\pm$1.30} & 5 & 1-b,b,a,a & \mysplit{19.99$\pm$1.30 \\ - \\ 17.24$\pm$1.30 \\ 16.51$\pm$1.30 \\ 11.74$\pm$1.30 } & \mysplit{ - \\ - \\ 36.73$\pm$1.30 \\ 21.12$\pm$1.30 } \\ 
  &              &        &       &                 &                    &   &       &                  &                    &   \\         
89 & J1835$-$1020 & 0.30 &   - & - & \mysplit{ - } & \mysplit{ - } & 1 & 1-a & \mysplit{10.80$\pm$0.70} & \mysplit{ - } \\ 
  &              &        &       &                 &                    &   &       &                  &                    &   \\         
90 & J1835$-$1106 & 0.17 &   1 & 1-a & \mysplit{26.70$\pm$1.30 } & \mysplit{ - } & 1 & 1-a & \mysplit{13.30$\pm$1.30 } & \mysplit{ - } \\ 
  &              &        &       &                 &                    &   &       &                  &                    &   \\         
91 & B1839+09 & 0.38 &  2 & 1/3-a,b & \mysplit{3.97$\pm$0.33 \\ 4.61$\pm$0.33 } & \mysplit{4.31$\pm$0.33 } & 2 & 1-a & \mysplit{ - \\ 6.40$\pm$0.33 } & \mysplit{ - } \\ 
  &              &        &       &                 &                    &   &       &                  &                    &   \\         
92 & B1839$-$04 & 1.84  & 2 & 1-a,a & \mysplit{45.48$\pm$ 2.08 \\ 34.80$\pm$ 2.08 } & \mysplit{55.16$\pm$0.20 } & 2 & 1-a,a & \mysplit{18.68$\pm$0.20 \\ 10.22$\pm$0.20 } & \mysplit{53.26$\pm$0.20 } \\ 
  &              &        &       &                 &                    &   &       &                  &                    &   \\         
93 & J1843$-$0000 & 0.88  & - & - & \mysplit{ - } & \mysplit{ - } & 1 & 1-a & \mysplit{12.23$\pm$0.14 } & \mysplit{ - } \\ 
  &              &        &       &                 &                    &   &       &                  &                    &   \\         
94 & B1842+14 & 0.38 &  2 & 1-a & \mysplit{8.24$\pm$0.33 } & \mysplit{ - } & 1/2 & 1-a & \mysplit{9.30$\pm$0.33 } & \mysplit{ - } \\ 
  &              &        &       &                 &                    &   &       &                  &                    &   \\         
95 & B1844$-$04 & 0.60 &   - & - & \mysplit{ - } & \mysplit{ - } & 2 & 1-b,b & \mysplit{8.44$\pm$0.21 \\ 6.41$\pm$0.21 } & \mysplit{5.48$\pm$0.21 } \\ 
  &              &        &       &                 &                    &   &       &                  &                    &   \\         
96 & B1845$-$01 & 0.66 &  - & - & \mysplit{ - } & \mysplit{ - } & - & - & \mysplit{ - } & \mysplit{ - } \\ 
  &              &        &       &                 &                    &   &       &                  &                    &   \\         
97 & J1848$-$1414 & 0.30 &  1 & 1-a & \mysplit{19.03$\pm$ 3.70 } & \mysplit{ - } & 1 & 1-a & \mysplit{13.4$\pm$ 3.70 } & \mysplit{ - } \\ 
  &              &        &       &                 &                    &   &       &                  &                    &   \\         
98 & B1846$-$06 & 1.45 &  - & - & \mysplit{ - } & \mysplit{ - } & 3 & 1-a & \mysplit{ - \\ 3.30$\pm$0.09 \\ - } & \mysplit{ - } \\ 
  &              &        &       &                 &                    &   &       &                  &                    &   \\         
99 & J1852$-$2610 & 0.34 &  2 & 1-a,a & \mysplit{5.80$\pm$0.38 \\ 6.88$\pm$0.37 } & \mysplit{19.75$\pm$0.37 } & 2 & 1-b,a & \mysplit{8.60$\pm$1.16 \\ 7.42$\pm$1.16 } & \mysplit{17.41$\pm$1.16 } \\ 
  &              &        &       &                 &                    &   &       &                  &                    &   \\         
100 & B1857$-$26 & 0.61  & 5 & 1-b,c,a,c,a & \mysplit{6.28$\pm$0.20 \\ 11.24$\pm$0.20 \\ 6.04$\pm$0.20 \\ 5.40$\pm$0.20 \\ 3.92$\pm$0.20 } & \mysplit{10.25$\pm$0.20 \\ 8.11$\pm$0.20 \\ 8.53$\pm$0.20 \\ 6.65$\pm$0.20} & 5 & 1-b,c,c,b & \mysplit{9.73$\pm$0.20 \\- \\ 8.44$\pm$0.20 \\ 8.12$\pm$0.20 \\ 4.24$\pm$0.20 } & \mysplit{ - } \\ 
  &              &        &       &                 &                    &   &       &                  &                    &   \\         
101 & J1901$-$0906 & 1.78  & 2 & 1-a,a & \mysplit{2.48$\pm$0.07 \\ 1.55$\pm$0.07} & \mysplit{9.29$\pm$0.07 } & 2 & 1-a,a & \mysplit{2.05$\pm$0.07 \\ 1.55$\pm$0.07	} & \mysplit{8.49$\pm$0.07 } \\ 
  &              &        &       &                 &                    &   &       &                  &                    &   \\         
102 & B1907+10 & 0.28 &   3 & 1-a & \mysplit{ - \\ 6.19$\pm$0.44 \\ - } & \mysplit{ - } & 3 & 1-c,a & \mysplit{5.48$\pm$0.47 \\ 4.80$\pm$0.44 \\ - } & \mysplit{9.05$\pm$0.44 } \\ 
  &              &        &       &                 &                    &   &       &                  &                    &   \\         
103 & B1907+03 & 2.33 &  3 & 1-c,a,a & \mysplit{17.07$\pm$0.27 \\ 33.93$\pm$0.27 \\ 7.43$\pm$0.27 } & \mysplit{25.45$\pm$0.27 \\ 32.11$\pm$0.27 } & 3 & 1-b,c,a & \mysplit{16.38$\pm$0.27 \\ 35.67$\pm$0.27 \\ 7.79$\pm$0.27 } & \mysplit{23.1$\pm$0.27 \\ 31.17$\pm$0.27  } \\ 
  &              &        &       &                 &                    &   &       &                  &                    &   \\         
104 & B1911$-$04 & 0.83  & 2 & 1/3-a & \mysplit{ - \\ 2.98$\pm$0.15 } & \mysplit{3.35$\pm$0.15 } & 2 & 1/3-a & \mysplit{ - \\ 3.21$\pm$0.15 } & \mysplit{3.44$\pm$0.15 } \\ 
  &              &        &       &                 &                    &   &       &                  &                    &   \\         
105 & B1914+09 & 0.27 &   2 & 1-a,b & \mysplit{6.18$\pm$0.46 \\ 4.91$\pm$0.47 } & \mysplit{7.21$\pm$0.46 } & 2 & 1-a,a & \mysplit{5.42$\pm$0.47 \\ 6.63$\pm$0.47} & \mysplit{8.2$\pm$0.46 } \\ 
  &              &        &       &                 &                    &   &       &                  &                    &   \\         
106 & B1915+13 & 0.19 &   2 & 1/3-a & \mysplit{5.22$\pm$0.64 \\ - } & \mysplit{ - } & 2 & 1/3-a & \mysplit{4.75$\pm$0.65 \\ - } & \mysplit{ - } \\ 
  &              &        &       &                 &                    &   &       &                  &                    &   \\         
107 & B1917+00 & 1.27 &  3 & 1-b,a & \mysplit{3.13$\pm$0.10 \\ 2.26$\pm$0.10 \\	- } & \mysplit{3.34$\pm$0.10 \\ 3.06$\pm$0.10 } & 3 & 1-b,a,b & \mysplit{2.67$\pm$0.10 \\ 1.97$\pm$0.10 \\ 2.45$\pm$0.10 } & \mysplit{3.27$\pm$0.10 \\ 3.62$\pm$0.10} \\ 
  &              &        &       &                 &                    &   &       &                  &                    &   \\         
108 & J1919+0134 & 1.60 &  - & - & \mysplit{ - } & \mysplit{ - } & 2 & 1-a,a & \mysplit{9.17$\pm$0.24 \\ 4.79$\pm$0.24 } & \mysplit{10.10$\pm$0.23 } \\ 
  &              &        &       &                 &                    &   &       &                  &                    &   \\         
109 & B1918+19 & 0.82 &  4 & 1-b,bc & \mysplit{5.42$\pm$0.46 \\ 12.37$\pm$0.46 \\ - \\ 14.21$\pm$0.46 } & \mysplit{14.24$\pm$0.46 } & 4 & 1-b,b & \mysplit{5.15$\pm$0.46 \\ 14.83$\pm$0.46 \\ - \\ - } & \mysplit{11.01$\pm$0.46 \\ - \\ -} \\ 
  &              &        &       &                 &                    &   &       &                  &                    &   \\         
110 & B1919+21 & 1.34 &  - & - & \mysplit{ - } & \mysplit{ - } & 3/2 & 1/2/3-b,b & \mysplit{2.29$\pm$0.09 \\ - \\ 3.41$\pm$0.09 } & \mysplit{ - } \\ 
  &              &        &       &                 &                    &   &       &                  &                    &   \\         
111 & B1929+10 & 0.23 &   2 & 1-a,b & \mysplit{5.32$\pm$0.55 \\ 6.18$\pm$0.55 } & \mysplit{5.09$\pm$0.55 } & 2 & 1-a,b & \mysplit{5.90$\pm$0.55 \\ 4.58$\pm$0.55 } & \mysplit{3.52$\pm$0.55 } \\ 
  &              &        &       &                 &                    &   &       &                  &                    &   \\         
112 & B1937$-$26 & 0.40  & 2 & 1-a,b & \mysplit{3.38$\pm$0.31 \\ 3.50$\pm$0.31 } & \mysplit{5.05$\pm$0.31} & 2 & 1-a,b & \mysplit{2.54$\pm$0.31 \\ 2.57$\pm$0.31 } & \mysplit{5.05$\pm$0.31 } \\ 
  &              &        &       &                 &                    &   &       &                  &                    &   \\         
113 & B1944+17 & 0.44 &  3 & 3-b/a & \mysplit{- \\ 7.89$\pm$0.28 \\ 6.37$\pm$0.28 } & \mysplit{ - \\ 11.98$\pm$0.28 } & 3 & 3-b/a & \mysplit{ - \\ 11.55$\pm$0.28 \\ 7.88$\pm$0.28} & \mysplit{9.51$\pm$0.28 \\ 10.85$\pm$0.28 } \\ 
  &              &        &       &                 &                    &   &       &                  &                    &   \\         
114 & B2003$-$08 & 0.58  & 5 & 1-b,b,a,b,b & \mysplit{8.70$\pm$0.65 \\ 6.73$\pm$0.65 \\ 8.77$\pm$0.65 \\ 8.2$\pm$0.65 \\ 8.17$\pm$0.65 } & \mysplit{13.28$\pm$0.65 \\ 16.03$\pm$0.65 \\ 14.20$\pm$0.65 \\ 10.53$\pm$0.65 } & 5 & 1-b,c,a,c,b & \mysplit{14.72$\pm$0.66 \\ 7.16$\pm$0.66 \\ 11.54$\pm$0.65 \\ 11.19$\pm$0.65 \\ 7.14$\pm$0.66 } & \mysplit{8.24$\pm$0.66 \\ 14.20$\pm$0.66 \\ 15.11$\pm$0.65 \\ 8.24$\pm$0.65 } \\ 
  &              &        &       &                 &                    &   &       &                  &                    &   \\         
115 & B2043$-$04 & 1.55  & 3 & 3-b /a & \mysplit{ - \\ 2.45$\pm$0.08 \\ 2.12$\pm$0.08 } & \mysplit{ - \\ 2.89$\pm$0.08 } & 3 & 3-b /a & \mysplit{ - \\ 1.80$\pm$0.08 \\ 1.49$\pm$0.08 } & \mysplit{ - \\ 2.86$\pm$0.08 } \\ 
  &              &        &       &                 &                    &   &       &                  &                    &   \\         
116 & B2044+15 & 1.14 &  - & - & \mysplit{ - } & \mysplit{ - } & 2 & 1-c,a & \mysplit{5.70$\pm$0.11 \\ 2.69$\pm$0.11 } & \mysplit{9.17$\pm$0.11 } \\ 
  &              &        &       &                 &                    &   &       &                  &                    &   \\         
117 & B2045$-$16 & 1.96  & 3 & 1-a,b,a & \mysplit{2.17$\pm$0.06 \\ 3.86$\pm$0.06 \\ 1.81$\pm$0.06 } & \mysplit{9.06$\pm$0.06 \\ 4.50$\pm$0.06 } & 3 & 1-a,b,a & \mysplit{2.67$\pm$0.06 \\ 4.45$\pm$0.06 \\ 1.74$\pm$0.06 } & \mysplit{8.52$\pm$0.06 \\ 3.96$\pm$0.06} \\ 
  &              &        &       &                 &                    &   &       &                  &                    &   \\         
118 & J2144$-$3933 & 8.51  & 1 & 1-a & \mysplit{0.80$\pm$0.06 } & \mysplit{ - } & 1 & 1-a & \mysplit{0.76$\pm$0.12 } & \mysplit{ - } \\ 
  &              &        &       &                 &                    &   &       &                  &                    &   \\         
119 & B2303+30 & 1.58 &  2 & 3-a/b & \mysplit{3.48$\pm$0.08 \\ 2.40$\pm$0.08 } & \mysplit{3.93$\pm$0.08 } & - & - & \mysplit{ - } & \mysplit{ - } \\ 
  &              &        &       &                 &                    &   &       &                  &                    &   \\         
120 & B2310+42 & 0.35 &  4 & 1-b,b,b & \mysplit{ - \\ 3.89$\pm$0.36 \\ 5.88$\pm$0.36 \\ 3.71$\pm$0.36 } & \mysplit{ - \\ 5.07$\pm$0.36 \\ 5.32$\pm$0.36 } & - & - & \mysplit{ - } & \mysplit{ - } \\ 
  &              &        &       &                 &                    &   &       &                  &                    &   \\         
121 & B2315+21 & 1.44 &  2 & 3-a/b & \mysplit{2.18$\pm$0.09 \\ 3.36$\pm$0.09 } & \mysplit{2.24$\pm$0.09 } & 2 & 3-a & \mysplit{2.05$\pm$0.26 \\ 2.54$\pm$0.26 } & \mysplit{2.39$\pm$0.26 } \\ 
  &              &        &       &                 &                    &   &       &                  &                    &   \\         
122 & B2327$-$20 & 1.64 &  3 & 1-a,a,b & \mysplit{1.44$\pm$0.08 \\ 1.86$\pm$0.08 \\ 2.64$\pm$0.08 } & \mysplit{2.70$\pm$0.08 \\ 1.84$\pm$0.08 } & 3 & 1-a,c & \mysplit{1.49$\pm$0.08 \\ 2.06$\pm$0.08 \\ - } & \mysplit{2.79$\pm$0.08 \\ 1.47$\pm$0.08} \\ 
  &              &        &       &                 &                    &   &       &                  &                    &   \\         
123 & J2346$-$0609 & 1.18 &  2 & 1-a,a & \mysplit{2.24$\pm$0.11 \\ 2.85$\pm$0.11 } & \mysplit{16.48$\pm$0.11 } & 2 & 1-a,a & \mysplit{2.27$\pm$0.21 \\ 3.23$\pm$0.21 } & \mysplit{14.04$\pm$0.21 } \\ 
  &              &        &       &                 &                    &   &       &                  &                    &   \\         

\end{longtable}



\normalsize
\section{\large \bf Quantile Regression} \label{sec:quant}
\normalsize
\noindent
The quantile regression is distinct from the traditional method of least 
squares. The least square finds the conditional mean function, $E(y|x)$, where 
$y$ corresponds to the mean value for the data at variable $x$. The quantile 
regression on the other hand gives a more generalized response function, 
$Q_q(y|x)$, where the quantile $q$ (0 $< q <$ 1) is such that $y$ splits the 
data at any $x$ with $q$ fraction below $y$ and 1-$q$ fraction above $y$. If 
the prediction error of a model function is $\epsilon_i$ for the $i^{th}$ 
variable, in least squares the minimization of the term $\sum_{i} \epsilon_i^2$
is carried out. In quantile regression asymmetric penalties are sought with 
weights (1-$q$)$|\epsilon_i|$ for over prediction and $q|\epsilon_i|$ for under
prediction. 

Let us consider $\hat{y}(x)$ to denote the prediction function and $\epsilon(x)
= y-\hat{y}(x)$ to denote the prediction error. Then $E(\epsilon(x)) = 
E(y-\hat{y}(x))$ denotes the loss associated with the prediction error. In the 
least square formulation the loss function is $E(\epsilon) = \epsilon^2$. 
In quantile regression if $\hat{y}^q(x)$ is the $q^{th}$ quantile prediction 
then the loss associated with the prediction error is asymmetric with either
$E_i^q(y-\hat{y}(x_i)) = q|y_i-\hat{y}(x_i)|$ if $y_i \geq \hat{y}(x_i)$ or
$E_i^q(y-\hat{y}(x_i)) = (1-q)|y_i-\hat{y}(x_i)|$ if $y_i < \hat{y}(x_i)$. The
objective function for minimization is given as :
\begin{equation}
Q(\hat{y}^q)=  \sum_{y_{i}\geq \hat{y}^q(x_i)} q|y_{i} - \hat{y}^q(x_i)| + \sum_{y_{i} < \hat{y}^q(x_i)} (1-q)|y_{i} - \hat{y}^q(x_i)|
\label{eq:quantile}
\end{equation}
The above function is non differentiable but can be minimized using the simplex
method to find the optimal solution for $\hat{y}^q(x)$. 

\begin{figure}[h]
\begin{center}
\begin{tabular}{@{}lr@{}}
{\mbox{\includegraphics[angle=0,scale=0.35]{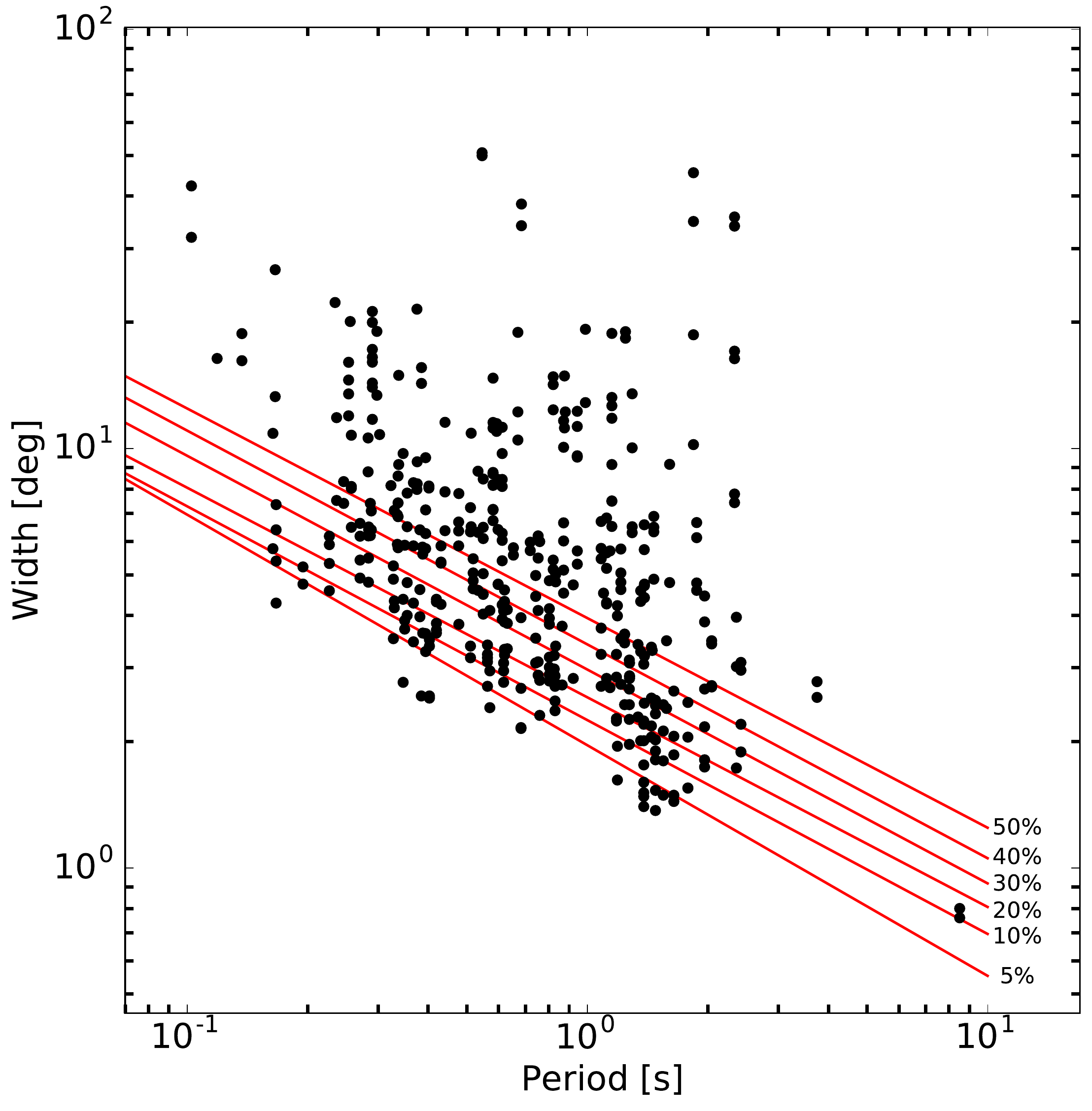}}}
\end{tabular}
\end{center}
\caption{The plot shows the quantile regression analysis for the component 
widths as a function of the period. The quantile regression is estimated for 
$q$ = 0.05, 0.1, 0.2, 0.3, 0.4, 0.5. The lower boundary has a period dependence
$\sim$ -0.5 for $q \geq$ 0.1, but the dependence is not well constrained for 
$q$ = 0.05. We have used $q$ = 0.1 as our estimate for the lower boundary which
correspond to $W = 2.25 P^{-0.51}$ in the above case.
\label{quant_reg}}
\end{figure}

The lower boundary line in the component width distribution with period was 
estimated using quantile regression. The data was converted to logarithmic 
scale in order to make the boundary line ($\hat{y}^q(x) = ax +b$) a linear 
function of period. The quantile regression is implemented in the statsmodel 
Python module's QuantReg class which was used for these estimates. We 
investigated the boundary line for multiple definition of the quantile $q$ = 
0.05, 0.1, 0.2, 0.3, 0.4, 0.5. The results for each case is shown in 
figure~\ref{quant_reg} where all the component widths have been considered. As 
seen in the figure the boundary line has a slope close to -0.5 for all $q \geq$ 
0.1 reproducing the period dependence. Hence, we have used $q$ = 0.1 as our 
quantile level for estimating the lower boundary line. 

\begin{thebibliography}{}

\bibitem[Arendt \& Eilek(2002)]{are02} Arendt, P.N.; Eilek, J.A.  2002, \apj, 581, 451
\bibitem[Asseo \& Melikidze(1998)]{ass98} Asseo, E.; Melikidze, G.I.  1998, \mnras, 301, 59
\bibitem[Bai \& Spitkovski(2010)]{bai10} Bai, X.N.; Spitkovski, A.  2010, \apj, 715, 1270
\bibitem[Basu \etal\ (2015)]{bas15} Basu, R.; Mitra, D.; Rankin, J.M.  2015, \apj, 798, 105
\bibitem[Basu \etal\ (2016)]{bas16} Basu, R.; Mitra, D.; Melikidze, G.I.; Maciesiak, K.; Skrzypczak, A.; Szary, A.  2016, \apj, 833, 29
\bibitem[Basu \etal\ (2017)]{bas17} Basu, R.; Mitra, D.; Melikidze, G.I.  2017, \apj, 846, 109
\bibitem[Beskin \etal\ (1993)]{bes93} Beskin, V.S.; Gurevich, A.V.; Istomin, Y.N.  1993, Physics of the pulsar magnetosphere, book.
\bibitem[Blaskiewicz \etal\ (1991)]{bla91} Blaskiewicz, M.; Cordes, J. M.; Wasserman, I.  1991, \apj, 370, 643
\bibitem[Chen \& Beloborodov(2014)]{che14} Chen, A.Y.; Beloborodov, A.M.  2014, \apjl, 795, L22
\bibitem[Craig \& Romani(2012)]{cra12} Craig, H.A.; Romani, R.W.  2012, \apj, 755, 137
\bibitem[Clemens \& Rosen(2004)]{cle04} Clemens, J.C.; Rosen, R.  2004, \apj, 609, 340
\bibitem[Dyks \& Harding(2004)]{dyk04} Dyks, J.; Harding, A.K.  2004, \apj, 614, 869
\bibitem[Dyks(2008)]{dyk08} Dyks, J.  2008, \mnras, 391, 859
\bibitem[Everett \& Weisberg (2001)]{eve01} Everett, J. E., Weisberg, J. M.  2001, \apj, 553, 341
\bibitem[Fung \etal\ (2006)]{fun06} Fung, P.K.; Khechinashvili, D.; Kuijpers J. 2006, \aap, 445, 779
\bibitem[Gil \etal\ (1984)]{gil84} Gil, J.A.; Gronkowski P.; Rudnicki W.  1984, \aap, 132, 312
\bibitem[Hakobyan \& Beskin(2014)]{hak14} Hakobyan, H.L.; Beskin, V.S.  2014, ARep, 58, 889
\bibitem[Hibschman \& Arons(2001)]{hib01} Hibschman, J.A.; Arons, J.  2001, \apj, 546, 382
\bibitem[Kijak \& Gil(1997)]{kij97} Kijak, J.; Gil, J.  1997, \mnras, 288, 631
\bibitem[Kijak \& Gil(2003)]{kij03} Kijak, J.; Gil, J.  2003, \aap, 397, 969
\bibitem[Krzeszowski \etal\ (2009)]{krz09} Krzeszowski, K.; Mitra, D.; Gupta, Y.; Kijak, J.; Gil, J.; Acharyya, A.  2009, \mnras, 393, 1617
\bibitem[Kumar \& Gangadhara(2012a)]{kum12a} Kumar, D.; Gangadhara, R.T.  2012a, \apj, 746, 157
\bibitem[Kumar \& Gangadhara(2012b)]{kum12b} Kumar, D.; Gangadhara, R.T.  2012b, \apj, 754, 55
\bibitem[Kumar \& Gangadhara(2013)]{kum13} Kumar, D.; Gangadhara, R.T.  2013, \apj, 769, 104
\bibitem[Maciesiak \etal\ (2011a)]{mac11a} Maciesiak, K.; Gil, J.; Ribeiro, V.A.R.M.  2011a, \mnras, 414, 1314
\bibitem[Maciesiak \etal\ (2011b)]{mac11b} Maciesiak, K.; Gil, J.  2011b, \mnras, 417, 1444
\bibitem[Maciesiak \etal\ (2012)]{mac12} Maciesiak, K.; Gil, J.; Melikidze, G.  2012, \mnras, 424, 1762
\bibitem[Melikidze \etal\ (2000)]{mel00} Melikidze, G.I.; Gil, J.A.; Pataraya, A.D.  2000, \apj, 544, 1081
\bibitem[Melikidze \etal\ (2014)]{mel14} Melikidze, G.I.; Mitra, D.; Gil, J.A.  2014, \apj, 794, 105
\bibitem[Melrose(2017)]{mel17} Melrose, D.B.  2017, RvMPP, 1, 5
\bibitem[Michel(1982)]{mic82} Michel, F.C.  1982, RvMP, 54, 1
\bibitem[Mitra \& Li(2004)]{mit04} Mitra, D.; Li, X.H.  2004, \aap, 421, 215
\bibitem[Mitra \& Seiradakis(2004)]{mit04b} Mitra, D.; Seiradakis, J.H. Hellenic Astronomical Society: Proceedings of the Sixth Astronomical Conference, held at Penteli, Athens, 15-17 September, 2003. Edited by Paul Laskarides. Published by the Editing Office of the University of Athens, Athens, Greece, 2004, p.205
\bibitem[Mitra \etal\ (2007)]{mit07} Mitra, D.; Rankin, J.M.; Gupta, Y.  2007, \mnras, 379, 932
\bibitem[Mitra \etal\ (2009)]{mit09} Mitra, D.; Gil, J.A.; Melikidze, G.I.  2009, \apjl, 696, L141
\bibitem[Mitra \& Rankin(2011)]{mit11} Mitra, D., Rankin, J.M.  2011, \apj, 727, 92
\bibitem[Mitra \etal\ (2016a)]{mit16a} Mitra, D.; Rankin, J.M.; Arjunwadkar, M.  2016a, \mnras, 460, 3063
\bibitem[Mitra \etal\ (2016b)]{mit16b} Mitra, D.; Basu, R.; Maciesiak, K.; Skrzypczak, A.; Melikidze, G.I.; Szary, A.; Krzeszowski, K.  2016b, \apj, 833, 28
\bibitem[Mitra(2017)]{mit17} Mitra, D.  2017, JApA, 38, 52
\bibitem[Petrova \& Lyubarskii(2000)]{pet00} Petrova, S.A.; Lyubarskii, Y.E.  2000, \mnras, 355, 1168
\bibitem[Philippov \& Spitkovsky(2014)]{phi14} Philippov, A.A.; Spitkovsky, A.  2014, \apjl, 785, L33
\bibitem[Philippov \etal\ (2015)]{phi15} Philippov, A.A.; Spitkovsky, A.; Cerutti, B.  2015, \apjl, 801, L19
\bibitem[Radhakrishnan \& Cooke(1969)]{rad69} Radhakrishnan, V.; Cooke, D.J.  1969, \aplett, 3, 225
\bibitem[Rankin(1990)]{ran90} Rankin, J.M.  1990, \apj, 352, 247
\bibitem[Rankin(1993a)]{ran93a} Rankin, J.M.  1993a, \apj, 405, 285
\bibitem[Rankin(1993b)]{ran93b} Rankin, J.M.  1993b, \apjs, 85, 145
\bibitem[Ruderman \& Sutherland(1975)]{rud75} Ruderman, M.A.; Sutherland, P.G.  1975, \apj, 196, 51
\bibitem[Smith \etal\ (2013)]{smi13} Smith, E.; Rankin, J.M.; Mitra, D.  2013, \mnras, 435, 1984
\bibitem[Szary \etal\ (2015)]{sza15} Szary, A., Melikidze, G.I., Gil, J.  2015, \mnras, 447, 2295 
\bibitem[Timokhin \& Arons(2013)]{tim13} Timokhin, A.N.; Arons, J.  2013, \mnras, 429, 20 
\bibitem[von Hoensbroech \& Xilouris(1997)]{von97} von Hoensbroech, A.; Xilouris, K.M.  1997, \aap, 324, 981 
\bibitem[Weltevrede \& Johnston(2008)]{wel08} Weltevrede, P.; Johnston, S.  2008, \mnras, 391, 1210
\end{thebibliography}
\end{document}